\documentclass{tlp}
\usepackage{latexsym}

\newtheorem{tho}{Theorem}[section]    
\newtheorem{de}{Definition}[section]
\newtheorem{lemm}[tho]{Lemma} 
\newtheorem{coro}[tho]{Corollary}
\newtheorem{ex}{Example}[section]

\newcommand{\fde}{$\hfill \Box$}
\newcommand{\La}{\Rightarrow}
\newcommand{\nat}{{\rm I\!N}}
  
\newcommand{\vlargo}{{|\!\mbox{---}}}
\newcommand{\vded}{{\|\!\mbox{---}}}
\newcommand{\Cequiv}{\vdash\!\dashv_{\cal C}}
\newcommand{\Requiv}{\vdash\!\dashv_{\cal R}}

\renewcommand{\thefootnote}{\fnsymbol{footnote}}

\title[CLP with Hereditary Harrop Formulas]
       {Constraint Logic Programming with \\ 
        Hereditary Harrop Formulas}
		
\author[J.Leach, S.Nieva, M.Rodr\'{\i}guez-Artalejo]
       {Javier Leach, Susana Nieva, Mario Rodr\'{\i}guez-Artalejo
        \thanks{This is a substantially revised and 
extended version of \protect\cite{ILPS}. The authors have been partially 
supported by the Spanish National Project TIC 98-0445-C03-02 {\em 
TREND} and the 
Esprit BRA Working Group EP-22457 {\em CCLII.}}\\
	    Dpto. Sistemas Inform\'{a}ticos y Programaci\'{o}n\\
        Av. Complutense s/n, Universidad Complutense de Madrid, \\ 
        E-28040 Madrid, Spain\\
\email{\{leach,nieva,mario\}@sip.ucm.es}}

\begin{document}
\maketitle  

\begin{abstract}
 Constraint Logic Programming ({\em CLP}\/) and Hereditary 
Harrop Formulas 
({\em HH}\/)are two well known ways to enhance the expressivity of 
Horn clauses. In
this paper, we present a novel combination of these two approaches. 
We show 
how to enrich the syntax and proof theory of {\em HH}\/ with the help of a
given constraint system, in such a way that
the key property of {\em HH}\/ as a logic programming language (namely, 
the existence of uniform proofs) is preserved. We also present a procedure
for goal solving, showing its soundness and completeness for 
computing answer constraints. As a consequence of this result, we 
obtain a new strong completeness theorem for {\em CLP}\/ that avoids 
the need to
build disjunctions of computed answers, as well as a more abstract
formulation of a known completeness theorem for {\em HH}. \\

\noindent{\bf keywords:}
 constraint systems, hereditary Harrop 
formulas, uniform proofs, goal solving.

\end{abstract}

\renewcommand{\thefootnote}{\arabic{footnote}}

\setcounter{footnote}{0}

 
\section{Introduction} \label{Intro}

\vspace{-2mm}
Traditionally, the logic of Horn clauses has been considered as the 
basis for logic programming \cite{vEK}. In spite of its Turing 
completeness \cite{AN 78}, the lack of expressivity of Horn clauses 
for programming purposes is widely acknowledged. During the last 
decade, different extensions of Horn clauses have been proposed, with 
the aim of increasing expressivity without sacrificing the 
declarative 
character of pure logic programming. Among such extensions, two 
important
approaches are Constraint Logic Programming ({\em CLP}\/) and 
Hereditary 
Harrop Formulas ({\em HH}\/). 

The {\em CLP}\/ scheme \cite{clp} goes beyond the limitations of
the Herbrand universe by providing the ability to program with Horn 
clauses over different computation domains, whose logical behaviour
is given by  constraint systems. {\em CLP}\/ languages keep all 
the good 
semantic properties of pure logic programming, including soundness 
and 
completeness results \cite{clprep}. Their implementation relies
on the combination of {\em SLD}\/ resolution with dedicated algorithms for 
constraint 
entailment, solving and simplification. Therefore, efficient and yet 
declarative 
programs can be written to solve complex combinatorial problems. See 
\cite{surv} 
for a survey of the foundations, implementation issues and 
applications of {\em CLP}\/ 
languages.

On the other hand, the {\em HH}\/ approach \cite{MNS87} overcomes 
the inability of Horn clauses to provide a logical basis for several 
constructions commonly found in modern programming languages, such as 
scoping, abstraction and modularity. This is achieved by extending 
Horn clauses to a richer fragment of intuitionistic logic that allows 
us to use disjunctions, implications and quantifiers in goals. 
In fact, {\em HH}\/ is a typical example of an {\em abstract logic 
programming 
language}, in the sense of \cite{Mil-al}. Abstract logic programming 
languages 
are characterized by the fact that the declarative meaning of a 
program, 
given by provability in a deduction system, can be interpreted 
operationally as 
goal-oriented search for solutions. Technically, the existence of 
{\em uniform
proofs}\/ for all provable goal formulas permits the search 
interpretation of 
provability. The implementation of programming languages based 
on {\em HH}, such as $\lambda$-Prolog \cite{MN,NM87}, 
requires the resolution of  the problem of unifying terms 
occurring under the scope of arbitrary quantifier prefixes. Correct 
unification 
algorithms for such problems have been studied in \cite{Mil92,Nad93}.
Moreover, \cite{Nad93} shows in detail the soundness and completeness 
of a goal 
solving procedure for the first-order {\em HH}\/ language.

The aim of this paper is to present a framework for the 
combination of the {\em CLP}\/ and {\em HH}\/ approaches, that 
incorporates the benefits of expressivity and efficiency that {\em HH}\/ 
and {\em CLP}\/ bring to logic programming, respectively.
We will enrich the 
syntax of first-order {\em HH}\/ with constraints coming from a given
constraint system. The resulting language is such 
 that all constructions and results  are valid for 
 any constraint system ${\cal C}$,
therefore  we can speak of a {\em scheme
HH}($X$) with {\em instances HH}(${\cal C}$), as in {\em CLP}.
We will define an amalgamated proof system that combines 
inference rules from intuitionistic sequent calculus with constraint entailment, 
in such a way that the key 
property of an abstract logic programming language  is preserved. 
Moreover, we will also present a sound and 
complete procedure for goal solving.
As in {\em CLP}\/, the result of solving a goal using a 
program will be an {\em answer constraint}.

The following simple program $\Delta$, goal $G$ and constraint $R$
 belong to the instance  
{\em HH}(${\cal R}$) given by the constraint system 
${\cal R}$ for real numbers. 
We will refer to this as  the {\em disc example}\/ in the sequel.
\begin{center}
	$\begin{array}{l}
\Delta \equiv \{\forall x \forall y ( x^2 + y^2 \leq 1 \Rightarrow disc \ 
(x,y))\}\\
G \equiv \forall y ( y^2 \leq 1/2 \Rightarrow disc \ (x,y))\\
R \equiv x^2 \leq 1/2
\end{array}$
\end{center}
In the  example,  the formula $R$ turns out to be 
a correct and computable answer constraint in the resolution of $G$ 
from $\Delta$. Due to the soundness and completeness of the goal solving
procedure, $G$ can be deduced from $\Delta$ and $R$ in the amalgamated
proof system.
In Figure 1 a uniform proof is presented   
of the sequent $\Delta; R \, \vlargo \, G$, using the inferences 
rules of the calculus ${\cal UC}$ which will be presented in Section 
\ref{Prove}.

\begin{figure}
\fbox{\begin{minipage}{11.5cm}
\begin{tabular}{cl}
$ x^2 \leq 1/2,  y^2 \leq 1/2  \, \vdash_{\cal R} \, 
	\exists u \exists v (x \approx u \wedge y \approx v
	 \wedge  u^2 + v^2 \leq 1)$ & \\ 
\rule{9cm}{0.1mm}  & $(C_{R})$  \\[2mm]
	$\Delta; x^2 \leq 1/2,  y^2 \leq 1/2  \, \vlargo \, 
	\exists u \exists v (x \approx u \wedge y \approx v
	 \wedge  u^2 + v^2 \leq 1)$ & \\ 
\rule{9cm}{0.1mm}  & $(Clause)$ \\[2mm]
	$\Delta; x^2 \leq 1/2,  y^2 \leq 1/2  \, \vlargo \, 
	 disc \ (x,y)$ & \\ 
	\rule{6cm}{0.1mm}  & $(\Rightarrow  \! C_{R})$  \\[2mm]
$\Delta; x^2 \leq 1/2 \, \vlargo \, 
	 y^2 \leq 1/2 \Rightarrow disc \ (x,y)$ & \\ 
		\rule{6.3cm}{0.1mm}  & $(\forall_{R})$ \\[2mm]
	$\Delta; x^2 \leq 1/2 \, \vlargo \, 
	\forall y ( y^2 \leq 1/2 \Rightarrow disc \ (x,y))$ &   
\end{tabular}
\end{minipage}}
\caption{}
\end{figure}
  
From a technical point of view, for the particular case of the 
 Herbrand 
constraint system, our completeness result boils down to a more 
abstract 
formulation of the completeness theorem in \cite{Nad93}. In the case 
of {\em CLP}\/ 
programs using only Horn clauses with constraints, our goal solving 
procedure 
reduces to constrained resolution, and our completeness theorem 
yields a form of
strong completeness for success that avoids the need to build 
disjunctions of computed 
answers, in contrast to \cite{Mah87}, Th. 2 (see also \cite{clprep}, 
Th. 4.12).
The reason for this discrepancy is that our amalgamated proof system 
uses more
constructive inference mechanisms to deduce goals from program 
clauses, as we will see.

The rest of this paper is organized as follows:
Section \ref{Exampl}
shows some programming examples, that illustrate the specific 
benefits of the combination of {\em CLP\/} and {\em HH\/}. In Section 
\ref{Syntax} we recall 
the notion of a constraint system and we define the syntax of {\em 
HH}\/ with 
constraints. In Section \ref{Prove} we present an intuitionistic proof 
system for 
{\em HH}\/ with constraints, and we show the existence of  uniform 
proofs, then  an equivalent proof system allowing only
uniform proofs is defined. Based on this second calculus,  
a sound and complete procedure for goal solving is presented as a 
transformation system in Section \ref{Solve}. In Section \ref{Concl}
we summarize conclusions and possible lines for future research.   
 In order to improve readability of the paper, some	
 proofs have	been omitted or compressed in the main text. Full proofs 
 appear  in  the Appendix.	

\section{Examples}\label{Exampl}
Although simple, the programs of this section exemplify the
 programming style in {\em HH}($X$) languages, combining the 
characteristic utilities of {\it HH} --such as to add 
temporarily facts to the program or to limit the scope of the names-- 
with the advantages of using constraint solvers, 
instead of syntactical unification. The syntax used in the examples is 
basically that of {\em HH}\/ languages, with the addition of 
constraints in clause bodies and goals. In particular, the notation 
$t \approx t'$ will be used for equality constraints. More formal 
explanations will follow in Section \ref{Syntax}. 

The programs below are based on a constraint system 
which is defined as a combination   of  ${\cal R}$
 (real numbers)
 and ${\cal H}$ (Herbrand universe). This constraint
 system underlies the well known language {\em CLP}(${\cal R}$) 
 \cite{clp(r)}. The elements in the intended computation domain can be 
 represented as trees whose internal nodes are labeled by 
 constructors,
  and whose leaves are labeled either by constant 
 constructors or by real numbers. In particular this includes the 
 representation of lists, possibly with real numbers as members.
 We will use Prolog's syntax for the list constructors.
 

\begin{ex}[Hypothetical queries in a data base system]  
 The following program keeps record of the marks of different 
students in two exercises they have to do to pass an exam.
\begin{center}
\fbox{\begin{minipage}{11cm}
\begin{tabular}{l}
{\it exercise$1(bob, 4)$.}\\
{\it exercise$1(fran, 3)$.}\\
{\it exercise$2(fran, 6)$.}\\
{\it exercise$1(pep, 5)$.}\\
{\it exercise$2(pep, 6)$.}\\[3mm]
{\it pass$(X) \Leftarrow$ exercise$1(X, N1) \wedge$exercise$2(X, N2) 
\wedge (N1 + N2)/2 >  5$.\frenchspacing}\\
\end{tabular}\end{minipage}
}
\end{center}

\noindent While the goal $G \equiv$ {\it pass$(bob)$}
 fails, $G' \equiv$ {\it exercise$2(bob, 6.5) \Rightarrow$ 
pass$(bob)$} succeeds. 
 To resolve this last goal, the fact  
{\it  exercise$2(bob, 6.5)$} is added to the program, but not permanently. 
If we  would put again the query $G \equiv$
{\it pass$(bob)$} it would fail again.

Suppose now we want to know the requirements a student has to fulfil
 to pass, then we add to the program the clauses:

\begin{center}
\fbox{\begin{minipage}{11cm}
\begin{tabular}{l}
{\it need-to-pass$(A, [])\Leftarrow$ pass$(A)$.}\\
{\it need-to-pass$(A, [ex1(X)|L]) \Leftarrow
($exercise$1(A, X) \Rightarrow$ need-to-pass$(A, L))$.}\\
{\it need-to-pass$(A, [ex2(X)|L]) \Leftarrow
($exercise$2(A, X) \Rightarrow$ need-to-pass$(A, L))$.}\\
\end{tabular}
\end{minipage}}
\end{center}

\noindent The goal $G \equiv$ 
{\it   need-to-pass$(bob, L)$}
 will produce an answer  equivalent in the constraint system to  
$\exists N (L \approx [ex2(N)]  \wedge  N > 6)$.

\noindent To get this answer, the intermediate goal 
{\it exercise$2(A, X) \Rightarrow$ need-to-pass$(A, L1)$}   
should be solved
with the constraint  $A \approx bob$. 
 This would require:

i) To introduce the fact 
{\it exercise$2(A, X)$} in the base.
Note that the effect is different to adding a clause 
in Prolog with {\it assert}, since this implies the universal 
quantification of 
{\it A} and {\it X}.
 
ii) Try to solve the goal {\it need-to-pass$(A, [])$} with the first 
clause of this predicate, so to solving 
{\it pass$(A)$}, with the constraint  $A \approx bob$ and  $L1 \approx 
[]$.
This will add the constraints
$X \approx N, (4 + N)/2 > 5$.

A similar example is shown in \cite{Hod},  here the benefit is in the 
use of constraints allowing  to write conditions about the real 
numbers that help to solve the goal more efficiently. \fde
\end{ex}

\begin{ex}[Fibonacci numbers]
\cite{Coh} uses the computation of Fibonacci numbers as a simple 
example to illustrate the advantages of constraint solving w.r.t. 
built-in arithmetic (as available in Prolog). The recursive definition 
of Fibonacci sequence gives rise immediately to the following 
{\em CLP}(${\cal R}$) program:
\begin{center}
\begin{tabular}{l}
{\it fib$(0,1)$.}\\
{\it fib$(1,1)$.}\\
{\it fib$(N,F1 + F2) \Leftarrow$
 $N \geq 2$ $\wedge$ fib$(N-1, F1))$ $\wedge$ fib$(N - 2, F2)$.}
\end{tabular}
\end{center}
Thanks to the abilities of the constraint solver, this program is 
reversible. In addition to goals such as {\it fib$(10, X)$}, with 
answer $X \approx 89$, we can also solve goals as {\it fib$(N, 89)$}
with answer $N \approx 10$. However, the program is based on an 
extremely inefficient double recursion. As a consequence, it runs in 
exponential time, and multiple recomputations of the same Fibonacci 
number occur.

In {\em HH}(${\cal R}$) we can avoid this problem by using 
implications in goals to achieve the effect of tabulation. At the 
same time, the program remains reversible and close to the 
mathematical specification of the Fibonacci sequence.
\begin{center}
\fbox{\begin{minipage}{11cm}
\begin{tabular}{l}
{\it fib$(N,X) \Leftarrow$ $($memfib$(0, 1) \Rightarrow ($memfib$(1, 1) 
\Rightarrow$  getfib$(N, X, 1)))$.}\\[-1mm]
{\it getfib$(N, X, M) \Leftarrow$ $0 \leq N$ $\wedge$ $N \leq M$
  $\wedge$ memfib$(N, X)$}.\\
 {\it getfib$(N, X, M) \Leftarrow$
 $N > M$ $\wedge$  memfib$(M-1, F1)$ $\wedge$ memfib$(M, F2)$ $\wedge$}\\
\hspace*{2.8cm} $(${\it memfib$(M +1, F1 + F2) \Rightarrow$ 
getfib$(N, X, M + 1))$.}\\
\end{tabular}
\end{minipage}}
\end{center}
A predicate call of the form {\it getfib$(N,X,M)$} assumes that the 
Fibonacci numbers {\it fib$_{i}$}, with $0 \leq i \leq M$, 
are memorized as atomic clauses for {\it memfib} in the local program. 
The call computes
the  $N$-th Fibonacci number in $X$; at the same time, the 
Fibonacci numbers {\it fib$_{i}$}, with $M < i \leq N$
are memorized during the computation.

Let us consider two simple goals for this program: 

i) $G_{1} \equiv fib(2,X)$. In order to solve $G_{1}$, 
{\it memfib$(0,1)$} and {\it memfib$(1,1)$} are added to the 
local program, and the goal {\it getfib$(2,X,1)$} is solved. Since 
$2>1$, the first clause for {\it getfib} fails. The second clause  
 for $\mbox{\it getfib}$ puts {\it memfib$(2,2)$} into the local program 
 and produces the new goal {\it getfib$(2,X,2)$}, 
 which is solved with answer $X \approx 2$ by means of the first clause.   

 ii) $G_{2} \equiv fib(N,2)$. Analogously, $G_{2}$ is solved by 
solving {\it getfib$(N,2,1)$} after adding 
{\it memfib$(0,1)$} and {\it memfib$(1,1)$} into the local program.
The   first clause for {\it getfib}  fails.
 Therefore, the constraint $N > 1$ is assumed and
the new goal {\it getfib$(N,2,2)$} must  be solved, after putting
the atom {\it memfib$(2,2)$} into the local program.
Now,  the first clause for {\it getfib} leads easily to
the  answer $N \approx 2$. 

\noindent In general, all goals of the two forms:

i)  {\it fib$(n,X)$}, $n$ given,

ii)  {\it fib$(N,f)$}, $f$ a given Fibonacci number

\noindent can be solved by our goal solving procedure. Moreover, goals 
of the form i) can be solved in $O(n)$ steps. In \cite{modu}, Miller 
showed that implicational goals can be used to store previously 
computed Fibonacci numbers, thus leading to an {\em HH}\/ 
 program that runs in time $O(n)$. 
Later  Hodas \shortcite{Hod} gave another memorized version 
of the computation of Fibonacci numbers,
 closer to the naive doubly recursive algorithm. Hodas' version 
combines implicational goals with a continuation-passing programming 
style which relies 
on  higher-order predicate variables. The benefit  
of our version w.r.t. \cite{modu,Hod} is the reversibility 
of the predicate {\em fib}\/ that is enabled by 
constraint solving.\fde 
\end{ex}

\begin{ex}[Relating some simple parameters in a 
mortgage]\label{mortage1}            
The following program  $\Delta$  is presented by Jaffar and 
Michaylov \shortcite{JM87}  
as an application of {\em CLP}(${\cal 
R}$).\footnote{This 
example is considered anew in \cite{clp(r)}.}
            
\begin{center}
\fbox{\begin{minipage}{11.5cm}
\begin{tabular}{l}
{\it mortgage$(P, T, I, M, B)\Leftarrow$}
{\it $0 \leq T \wedge T \leq 3 \ \wedge$ TotalInt} $\approx 
T*(P*I/1200) \wedge$\\
\hspace*{2cm}$B \approx P + {\it TotalInt} - (T*M)$.\\
{\it mortgage$(P, T, I, M, B)\Leftarrow$}
{\it $T > 3$ $\wedge$ QuartInt $\approx 3*(P*I/1200)\wedge$}\\
\hspace*{2cm} {\it 
mortgage}$(P+{\it QuartInt}-3*M,T-3, I, M, B)$.\\
\end{tabular} 
\end{minipage}}
\end{center}
Where
	 $P$ stands for principal Payment, 
            $T$ for Time in months, 
            $I$ for Interest rate, 
             $M$ for Monthly payment, and 
            $B$ for outstanding Balance. 
	
            In {\em CLP}(${\cal R}$) the goal $G \equiv {\it mortgage} (P, 6, 
10, M, 0)$,  produces the answer 
            $0 \approx 1.050625*P - 6.075*M$.
            From this answer we can deduce that  $P/(T*M) \approx P/(6*M) 
\approx 0.9637$ (the number 0.9637 is calculated as an approximation), 
            where
            $P/(T*M)$ represents the quotient of loss for delayed 
payment.

		 We consider now a more complicated problem, namely to find
{\it Imin}, {\it Imax} (with $0 \leq \mbox{\it Imin}\leq 
\mbox{\it Imax}$) such that any mortgage whose quotient of loss lies in 
the interval  [0.9637  . .  0.97] can be balanced in 6 months with some 
interest rate $I$ lying in the interval  $[${\it Imin} . . {\it 
Imax}$]$. This problem can be formulated in {\em HH}(${\cal R}$) by the goal:
             \begin{center}$G \equiv $
            $\forall M  \forall P ( 0.9637 \leq P/(6*M) \leq 0.97 
\Rightarrow$\\
            $\exists I ( 0 \leq {\it Imin} \leq I \leq {\it Imax} \wedge  
            mortgage(P, 6, I, M, 0))).$\end{center}
	Using the goal transformation rules {\it i)} -- {\it viii)} of Section 
	\ref{Solve}, we can show a   resolution of $G$ from $\Delta$
	that computes the answer constraint:
$${\it Imax} \approx 10 
            \wedge {\it Imin} \approx 8.219559  \mbox{ (approx.).}$$            
            More details on the resolution of this goal will be given 
            in Example \ref{mortage2} at the end of Section \ref{Solve}.  
\fde
\end{ex}

\section{Hereditary Harrop	Formulas with Constraints} \label{Syntax}

As explained in the Introduction, the framework presented in this paper requires 
the enrichement of the syntax of {\em Hereditary Harrop Formulas}\/ (shortly, 
{\em HH}\/) 
\cite{MNS87,Mil-al} with constraints coming from a given {\em constraint system}. Following \cite{Sar}, we view a constraint system as a pair ${\cal C} =
({\cal L}_{\cal C}, \vdash_{\cal C})$, where ${\cal L}_{\cal C}$ is the set of 
formulas allowed as constraints and 
$\vdash_{\cal C}$ $\subseteq$ ${\cal P}({\cal L}_{\cal C}) \times 
{\cal L}_{\cal C}$ is 
an {\em entailment relation}.  We use $C$ and $\Gamma$ to represent a constraint and a finite 
set of constraints, respectively. Therefore, $\Gamma \vdash_{\cal C} C$ means that 
the constraint $C$ is entailed by the set of constraints $\Gamma$.  We write 
just $\vdash_{\cal C} C$ if $\Gamma$ is empty. In 
\cite{Sar}, ${\cal L}_{\cal C}$ and $\vdash_{\cal C}$ are required to satisfy certain 
minimal assumptions, mainly related to the logical behaviour of $\wedge$ and 
$\exists$. Since we have to work with other logical symbols, our assumptions 
must be extended to account for their proper behaviour. Therefore, we assume: 

\begin{enumerate}

	\item[i)]  
	${\cal L}_{\cal C}$ is a set of formulas including $\top$ (true),
	$\bot$ (false) and  all the equations $t \approx t'$
	between terms over some fixed signature, and closed under 
	$\wedge, \Rightarrow,$ $ \exists, \forall$ and the application of
	substitutions of terms for variables.

	\item[ii)] 
	$\vdash_{\cal C}$ is {\em compact}, i.e., $\Gamma \vdash_{\cal C} C$
	holds iff $\Gamma_{0} \vdash_{\cal C} C$ for some finite $\Gamma_{0}
	\subseteq \Gamma$. $\vdash_{\cal C}$ is also {\em generic}, i.e., 
	$\Gamma \vdash_{\cal C} C$ implies $\Gamma \sigma \vdash_{\cal C} C \sigma$
	for every substitution $\sigma$.

	\item[iii)] 
	All the inference rules related to $\wedge, \Rightarrow, \exists, \forall$ 
	and $\approx$ valid in the intuitionistic fragment of first-order logic
	are also valid to infer entailments in the sense of $\vdash_{\cal C}$. 
	
\end{enumerate}

\noindent The notation $C \sigma$ used above means application to a constraint $C$ of a substitution
$\sigma = [t_{1}/x_{1},\ldots,t_{n}/x_{n}]$, using proper 
renaming of the variables bound in $C$ to avoid capturing free variables from the 
terms $t_{i}$, $1 \leq i \leq n$. 
$\Gamma \sigma$ represents the application of $\sigma$ to every 
constraint of the set $\Gamma$.
 In the sequel, the notation $F \sigma$ will also be used for 
other formulas $F$, not necessarily constraints.

Note that the three conditions i), ii), iii) are meant as minimal 
requirements. In particular, the availability of the equality symbol 
$\approx$ is granted in any constraint system, and it will always 
stand for a congruence. However, other specific axioms for equality 
may be different in different constraint systems. 

 Observe also that item iii) above, does not mean that $\vdash_{\cal C}$ is
restricted to represent deducibility in some intuitionistic theory.
 On the contrary, 
our assumptions allow us to consider constraint systems ${\cal C}$ such that 
${\cal L}_{\cal C}$ is a full first-order language with classical negation, and 
$\Gamma \vdash_{\cal C} C$ holds iff $Ax_{\cal C} \cup \Gamma \vdash C$, where
$Ax_{\cal C}$ is 
a suitable set of first-order axioms and $\vdash$ is the entailment relation of 
classical first-order logic with equality. In particular, three important 
constraint systems of this form are: ${\cal H}$, where $Ax_{\cal H}$ 
is Clark's axiomatization of the Herbrand universe \cite{Cla};
${\cal CFT}$, where $Ax_{\cal CFT}$ is Smolka and Treinen's axiomatization 
of the domain of {\em feature trees}\/ \cite{ST}; and ${\cal R}$, where $Ax_{\cal R}$ 
is Tarski's axiomatization of the real numbers \cite{Tar}. In these three cases, 
the constraint system is known to be {\em effective}, in the sense that 
the validity of entailments $\Gamma \vdash_{\cal C} C$, with finite $\Gamma$, 
can be decided by an effective procedure. 

The previous 
systems include the use of disjunctions. In {\em CLP\/} 
there is a well known completeness theorem due to Maher  
\shortcite{Mah87}, which relies on the possibility of building finite 
 disjunctions of  computed answer constraints. As we will see in 
 Section \ref{Solve}, 
 disjunctions are not needed in order to prove 
 completeness of goal solving in our setting. 
This is the reason why we do not enforce 
 ${\cal L}_{\cal C}$ to be closed under $\vee$ in the general case.

In the sequel, we assume an arbitrarily fixed effective constraint system 
${\cal C}$. By convention, the notation	
$\Gamma \vdash_{\cal C}\Gamma'$ will mean that $\Gamma \vdash_{\cal C} C$
holds for all $C \in \Gamma'$, and $C \Cequiv C'$ will abbreviate that $C 
\vdash_{\cal C} C'$ and $C'\vdash_{\cal C} C$ hold.
 Also, we will say that a constraint $C$ with free	
variables $x_{1}, \ldots, x_{n}$ is	${\cal C}$-satisfiable iff 
$\vdash_{\cal C} \,	\exists	x_{1}\ldots	\exists	x_{n} C$.

In order to define the syntax of the first-order formulas of 
{\em HH}(${\cal C}$), we assume a set 
$PS = \bigcup_{n \in \nat}PS^{n}$ of ranked 
predicate symbols (disjoint from the
symbols occurring in ${\cal L}_{\cal C}$) which are used to build atomic 
formulas $A$ of the form $P(t_1, \ldots, t_n)$,  with $P \in PS^{n}$.

\begin{de}
The set of
{\em definite clauses}, with elements noted $D$,
 and  the set of {\em goals}, with elements noted $G$, are defined by the following syntactic rules:
\begin{center}
$D :=   A \,|\, D_1 \wedge D_2 
\,|\,G  \Rightarrow  A\,|\,\forall x D$\\
$G := 
A \,|\, C \,|\, G_1 \wedge G_2 \,|\, G_1 \vee G_2 \,|\,
D  \Rightarrow  G  \,|\,C  \Rightarrow  G  \,|\,
\exists x G  \,|\,\forall x G$
\end{center}
\end{de} 

This syntax is the natural extension of first-order {\em HH}\/ as 
presented  in \cite{Nad93}. The novelty is that  
 constraints can occur in goals of the forms $C$ and $C \Rightarrow G$, 
and therefore also in definite clauses of the form $G \Rightarrow A$. 
Some variants could be considered, as e.g. dropping $D_1 \wedge D_2$ 
or replacing $G  \Rightarrow  A$ by $G  \Rightarrow  D$, but these 
changes would render a logically equivalent system. 
In the 
rest of the paper, by a {\em program}\/ we understand any finite set $\Delta$ of 
definite clauses. 
 This includes both {\em CLP}\/ programs and first-order 
 {\em HH}\/ programs as  particular cases.

As usual in the {\em HH}\/ framework, see e.g. \cite{Nad93}, we will work with
 a technical device (so-called {\em 
elaboration}\/) for decomposing the clauses of a given program into a simple 
form. This is useful for a natural formulation of goal solving 
procedures.

\begin{de}
We define the  {\em elaboration of a program}\/ $\Delta$ as the set 
$elab(\Delta) = 
 \bigcup_{D \in \Delta} elab(D),$  where $elab(D)$ is defined by 
case analysis in the following way:\\
  	-- $elab(A) =\{\top \Rightarrow A\}$.\\  	
  	-- $elab(D_1\wedge D_2) = elab(D_1) \cup elab(D_2)$.\\   
  	-- $elab(G\Rightarrow A) =\{G\Rightarrow A\}$. \\ 
   	-- $elab(\forall x D) =\{\forall x D' \, | \, D' \in elab(D)\}$.  
\end{de}
  
Note that all clauses in $elab(\Delta)$ have the form
  $\forall x_1\ldots \forall x_n (G \Rightarrow A),  n \geq 0.$  
  We still need another technicality.
A {\em 
  variant}\/ of such a clause is any clause of the form
   $\forall y_1\ldots \forall y_n (G\sigma \Rightarrow A\sigma)$
 where   $y_1,\ldots, y_n$ are new variables not occurring free in the 
 original clause, and $\sigma=[y_1/x_{1},\ldots, y_n/x_{n}]$.

\section{Proof Systems}\label{Prove}

In this section we  present an  amalgamated proof system ${\cal IC}$ 
that combines the usual inference rules from intuitionistic logic with 
the entailment relation $\vdash_{\cal C}$ of 
a constraint 
system ${\cal C}$.
We will derive  sequents of the form 
 $\Delta;\Gamma \,\vlargo \, G$
 where $\Delta$ is a program, $\Gamma$ represents 
 a finite set of 
constraints and $G$ is an arbitrary goal. We also show that ${\cal IC}$
enjoys completeness of uniform proofs, and we present a second proof 
system ${\cal UC}$ which is equivalent to ${\cal IC}$ in deductive 
power, but is tailored to build uniform proofs only.

 \subsection{The calculus ${\cal IC}$}
${\cal IC}$
stands for an \underline{I}ntuitionistic sequent calculus 
for {\em HH}(${\cal C}$) that allows to 
deduce a goal from defined clauses in the presence of  \underline{C}onstraints. 

The intuitionistic calculus with constraints $\vdash_{\cal IC}$ 
is defined as follows. 
 $\Delta;\Gamma \vdash_{\cal IC} \, G$ if and only if the sequent 
 $\Delta; \Gamma \,\vlargo \, G$ has a proof using the rules of the 
proof system ${\cal IC}$ that we introduce in the following.
 A proof of a 
sequent is a tree whose nodes are sequents, the root is the  sequent 
to be proved and the leaves 
match axioms of the calculus. The  rules  
regulate  relationship between child nodes and  parent nodes.
In the representation of the rules, we have added to the premises
the side conditions relating to the existence of proofs in the 
constraint system; these entailment relations are not considered as 
nodes of the proofs seen as trees. This notation simplifies the 
reading of both inference rules and proof trees.
\begin{itemize}
\item {\it Axioms to deal with  atomic goals or  
constraints}: 
$$
\frac{\Gamma \, \vdash_{\cal C} \, C}{\Delta; \Gamma \,\vlargo \, C}\;(C_{R})
\hspace{1 cm}
\frac{\Gamma \, \vdash_{\cal C} A\approx A' 
}{\Delta, A; \Gamma \,\vlargo \, A'}\;(Atom)
$$
 In $(Atom)$, $A, A'$  are assumed to begin with  the same predicate 
 symbol. $A \approx A'$ abbreviates  
 $t_1 \approx t'_1 \wedge \ldots \wedge t_n \approx t'_n$, where
  $ A \equiv  P(t_1, \ldots, t_n),$ $A' \equiv P(t'_1, \ldots, t'_n)$.
  
\item {\it Rules introducing the connectives and quantifiers 
of the Hereditary Harrop formulas}:
$$
\frac{\Delta;\Gamma \, \vlargo \,G_i}
      {\Delta;\Gamma \, \vlargo \,G_1  \vee  G_2}\;  (\vee_R)
 \;   (i=1,2) 
$$

$$
\frac{\Delta, D_1, D_2;\Gamma \, \vlargo \,G}
      {\Delta, D_1  \wedge  D_2 ; \Gamma\, \vlargo \,G}\;  (\wedge_L)  
 \hspace*{1cm} 
\frac{\Delta;\Gamma \, \vlargo \,G_1\;\;\;
     \Delta;\Gamma \, \vlargo \,G_2}
     {\Delta;\Gamma \, \vlargo \,G_1\wedge G_2}\; (\wedge_R)
$$

$$
\frac{\Delta;\Gamma \, \vlargo \,G_{1}\;\;\;
      \Delta,A;\Gamma \, \vlargo \,G}
  {\Delta,G_1 \Rightarrow A;\Gamma \, \vlargo \,G}\; 
(\Rightarrow_L)
$$

$$
\frac{\Delta,D;\Gamma \, \vlargo \,G}
     {\Delta;\Gamma \, \vlargo \,D \Rightarrow G}\; (\Rightarrow_R)
     \hspace*{1cm}
\frac{\Delta;\Gamma, C \, \vlargo \,G}
     {\Delta;\Gamma \, \vlargo \,C \Rightarrow G}\; (\Rightarrow \! 
     C_{R})      
$$

$$
\frac{\Delta;\Gamma, C \, \vlargo \,G[y/x] \;\;\;
\Gamma \, \vdash_{\cal C} \exists y C  }
  {\Delta;\Gamma \, \vlargo \,\exists x G}\; (\exists_R)
$$ 
\vspace{-6mm} 
$$ y \mbox{ does not appear free in the 
sequent of the conclusion.}$$

$$
\frac{\Delta,D[y/x];\Gamma, C \, \vlargo \,G\;\;\;
\Gamma \, \vdash_{\cal C} \exists y C  }
{\Delta,\forall x D;\Gamma \, \vlargo \,G}\; (\forall_L) 
\hspace*{1cm}
\frac{\Delta;\Gamma \, \vlargo \,G[y/x]}
{\Delta;\Gamma \, \vlargo 
\,\forall x G}\; (\forall_R)
$$
\vspace{-6mm} 
$$ \mbox{in both, }y\mbox{ does not appear free in the 
sequent of the conclusion.}$$
\end{itemize}

Note that the rule of contraction seems to be absent from this system, 
but in fact it is implicitly present because 
$\Delta$ and $\Gamma$ are viewed as sets 
(rather than sequences) in any sequent $\Delta;\Gamma \, \vlargo \,G$. 
In many respects, the inference rules of ${\cal UC}$
are similar to those used for {\em HH\/} in 
the literature; see e.g. \cite{Mil-al,Nad93}. However, the 
presence of constraints induces some modifications. Of particular 
importance are the modifications introduced to $(\exists_{R})$
and $(\forall_{L})$. 
A simple reformulation of the traditional
version of $(\exists_{R})$,
using a constraint $y \approx t$ instead of a substitution
$[t/x]$, representing an instance of $x$, could be: 
$$
\frac{\Delta;\Gamma, y\approx t \, \vlargo \,G[y/x]}
{\Delta;\Gamma \, \vlargo \,\exists x G}
$$
\vspace{-6mm}
$$\mbox{if $y$ does not 
occur in t, and it does not appear free in the conclusion.}$$

 In our constraint-oriented formulation of  
 $(\exists_{R})$  we allow any 
 satisfiable constraint $C$ (not necessary of the form $y \approx t$)
 instead of the substitution, in order to guess an instance of $x$.
 The next  example  shows that this 
 extra generality is necessary. 
   
   \begin{ex}\label{exis}
 This example is based on {\em HH}(${\cal R}$).
Consider

$\Delta \equiv \{\forall x  ( x^2\approx 2 \Rightarrow r (x))\}$,

$G \equiv \exists x \  r  (x)$.
	
\noindent The sequent $\Delta;  \, \vlargo \, G$ is expected to be 
derivable. However, the traditional formulation of $(\exists_{R})$ 
does not work, because no term $t$ in the language ${\cal L}_{\cal R}$ 
denotes a square root of 2. With our $(\exists_{R})$, choosing the 
${\cal R}$-satisfiable constraint $C \equiv x^2 \approx 2$, the 
problem is reduced to the easy derivation of the sequent
$\Delta; x^2 \approx 2 \, \vlargo \, r  (x)$. 
\fde
\end{ex}

Our definition
 of $(\forall_{L})$ is  dual to
 $(\exists_{R})$ and follows the same idea, since 
 $(\forall_{L})$ also relies on guessing an instance for $x$.
 On the other hand, rule $(\forall_{R})$ has a universal character.
 Therefore, the traditional formulation by means of a new variable 
 has been kept in this case.
  
 For technical reasons we need to measure the {\em size\/} of proofs. We
formalize this notion as the number of sequents in it, that coincides 
with the number of nodes of the proof seen as a tree.

In the sequel we will use some technical  properties of ${\cal 
IC}$-provability. Let us state them in the  next lemmas, whose 
proofs can be found in the Appendix.

The first lemma guarantees  that  
substitution of a term for a variable in a sequent, preserves  
${\cal IC}$-provability.

\begin{lemm}\label{sust}
 For any $\Delta, \Gamma,  G$, $x$ and $t$,
if $\Delta; \Gamma\, \vdash_{\cal IC}\, G$,
then there is  a proof of the same 
size of 
$\Delta[t/x]; \Gamma[t/x] \, \vlargo\, G[t/x]$. 
\end{lemm}
The next lemma  shows that 
a sequent continues to be provable if we 
strengthen the set of constraints.  

 \begin{lemm}\label{cons} For any $\Delta, \Gamma,  G$,
if 
$\Gamma'$ is a set of constraints such that
$\Gamma'\, \vdash_{\cal C}\, \Gamma$, and
$\Delta; \Gamma\, \vdash_{\cal IC}\, G$, then 
$\Delta;\Gamma'\, \vlargo\, G$ has a proof of the same size. 
\end{lemm}

\begin{coro}\label{ren}
 For any $\Delta, \Gamma,  G$, $x$ and $u$,
if $\Delta; \Gamma\, \vdash_{\cal IC}\, G$,
then 
$\Delta[u/x]; \Gamma, x \approx u \, \vlargo\, G[u/x]$ has a proof of the same 
size. 
\end{coro}
\begin{proof}
 By Lemma \ref{sust}, $\Delta[u/x]; \Gamma[u/x]
  \, \vlargo\, G[u/x]$ has a proof of the same size as 
  $\Delta; \Gamma \, \vlargo\, G$. Hence, applying Lemma \ref{cons},
  $\Delta[u/x]; \Gamma, x \approx u \, \vlargo\, G[u/x]$ has a proof of the same 
size, because $\Gamma, x \approx u \, \vdash_{\cal C}\, \Gamma[u/x]$.
 \end{proof}

The next lemma assures that free variables that appear only in the set of 
constraints of a sequent can be considered as existentially 
quantified in the proof of the sequent.
 
\begin{lemm}\label{exist}
 For any $\Delta, \Gamma, C, G$,
if $\Delta; \Gamma, C\, \vdash_{\cal IC}\, G$ and $x$ is a variable that 
does not appear free in $\Delta, \Gamma, G$,
then 
$\Delta; \Gamma, \exists x C\, \vlargo\, G$ has a proof of the same 
size. 
\end{lemm}

\subsection{Uniform proofs}

We are aiming at an {\em abstract logic programming 
language} in the sense of \cite{Mil-al}. This means that {\em uniform proofs}
must exist for all provable sequents. In our setting the idea of uniform proof 
consists in breaking down a goal into its components until obtaining an 
atomic formula or a constraint, before using the rules for 
introduction of connectives on the left or resorting to 
constraint entailment.

More formally, the notion of uniform proof is as follows.

\begin{de}
An ${\cal IC}$-proof is called {\em uniform proof}
when each  internal node in 
the proof tree is a sequent whose right-hand side $G$ is 
neither a constraint nor an atomic formula. Moreover  the inference 
rule   relating 
this node to its children must be  one of the right-introduction rules 
 $(\vee_R)$,
  $(\wedge_R)$,
  $(\Rightarrow_R)$,
  $(\Rightarrow  \! C_{R})$,      
  $(\exists_R)$, 
   $(\forall_R)$, 
according to the outermost logical symbol of $G$.
\end{de}

In order  to prove that uniform proofs exist for all 
${\cal IC}$-provable sequents, we follow the same approach that in 
\cite{Mil-al}, showing that any given ${\cal IC}$-proof can be 
transformed into a uniform proof. This is achieved by the next lemma.

\begin{lemm}[Proof Transformation] \label{trans}
 If $G$ is a goal, 
$\Delta$ a program and  $\Gamma$ a  set of constraint formulas, such that   
$\Delta;\Gamma \, \vlargo\, G$ has a proof of size $l$, then:
\begin{enumerate}
	\item  For $G \equiv A$,
		there are  $n$ constraint formulas $C_{1},\ldots,C_{n}$ ($n \geq 0$)
	and a formula $\forall x_{1}\ldots \forall x_{n}$ $(G' \Rightarrow A')$
	that is a variant of some formula in $elab(\Delta)$ such that
	$x_{1},\ldots\!,\! x_{n}$ are new distinct variables not appearing 
	free in $\Delta,\Gamma, A$, where 
	$x_{i}$ does not appear free in $C_{1}, \ldots\!,\! C_{i-1}$, for $1<  i 
\leq n$, 
	and $A'$ begins with the same 
	predicate symbol as $A$. In addition it holds: 
	\begin{enumerate}
		\item $\Gamma \, \vdash_{\cal C} \,\exists x_{1}
		C_{1}$; \hspace*{2mm}
		$\Gamma, C_{1} \, \vdash_{\cal C} \,\exists x_{2}
		C_{2}; \hspace*{2mm} \ldots ; \hspace*{2mm}
		\Gamma, C_{1}, \ldots ,C_{n-1} \, \vdash_{\cal C} \,\exists x_{n}
		C_{n}$.
	
		\item 
           $\Gamma, C_{1},\ldots,C_{n} \, \vdash_{\cal C} \, A'  \approx 
           A.$       
         \item $\Delta;\Gamma, C_{1},\ldots,C_{n} \, \vlargo\, 
         G'$
	 has a proof of size less than  $l$, or $G' \equiv \top$.
	
	\end{enumerate}
	
        \item If  $G \equiv C$, then  $\Gamma \, \vdash_{\cal C} \, C$.

	\item  If $G \equiv G_1\wedge G_2$, then  $\Delta;\Gamma \, 
	\vlargo\, G_1$ and $\Delta;\Gamma \, \vlargo\, G_2$ have proofs of 
size less than $l$.
 
	\item  If $G \equiv G_1\vee G_2$, then  $\Delta;\Gamma \, 
	\vlargo\, G_i$ has a proof of size less than $l$ for $i=1$ or $2$.

	\item  If $G \equiv D\Rightarrow G_1$, then $\Delta, D;\Gamma \, 
	\vlargo\, G_1$ has a proof of size less than $l$.  
	
		\item  If $G \equiv C\Rightarrow G_1$, then $\Delta;\Gamma, C \, 
	\vlargo\, G_1$ has a proof of size less than $l$.  

	\item For $G \equiv \exists x G_1$,  if $y$ is a variable not appearing 
	free in $\Delta,\Gamma, G$, then there is a constraint formula
	 $C$ such that:
	 \begin{enumerate}
	 \item $\Gamma \, \vdash_{\cal C} \,\exists y C$.
	 \item  
     $\Delta;\Gamma, C \, 
	\vlargo\, G_1[y/x]$ has a proof of size less than $l$. 
	\end{enumerate} 

	\item  If $G \equiv \forall x G_1$, then 
       $\Delta;\Gamma \, \vlargo\, $ $ G_1[y/x]$ has a proof of size 
less than  $l$, where $y$ is  a variable  that 
does not appear free in $\Delta, \Gamma, G$. 

\end{enumerate}
\end{lemm}
 \begin{proof} We reason by   
 induction on the size $l$ of the proof of 
$\Delta;\Gamma \, \vlargo\, G$, analyzing cases according to the last 
inference rule applied in the 
proof of the sequent $\Delta;\Gamma \, \vlargo\, G$.
A detailed proof can be found in the   Appendix. As novelties w.r.t. 
\cite{Mil-al}, we must deal with constraints and with the new 
formulation of rules $(\exists_{R})$, $(\forall_{L})$. 
Here we only sketch the case where $(\forall_{L})$
is the last inference rule applied and $G \equiv \exists w G_{1}$.
  Let us show graphically the proof transformation,
  in which we will 
 essentially switch the applications of $(\forall_{L})$ and 
 $(\exists_{R})$. By the induction hypothesis, the initial proof has the 
 form: \\
 
 \qquad	$\Delta', D[u/x]; \Gamma, C'\wedge C, u \approx y  \, \vlargo \, 
	G_{1}[z/w]$ 
	
\qquad\qquad \qquad $\uparrow$ Cor. \ref{ren}, Lem. \ref{cons} 
 
\qquad\qquad	$\Delta', D[y/x]; \Gamma, C', C  \, \vlargo \, 
	G_{1}[z/w]$ \qquad\qquad \qquad
	$\Gamma, C' \, \vdash_{\cal C}\,\exists z C$ 
\vspace{-3mm}  
\begin{center} 
\rule{10cm}{0.1mm} \ \   $(\exists_{R})$  
\end{center}
\vspace{-3mm}  
\qquad\qquad\qquad$\Delta', D[y/x]; \Gamma, C' \, \vlargo \, 
	\exists w G_{1} \qquad  \qquad\qquad\Gamma \,
	\vdash_{\cal C}\,\exists y C'$ 
\vspace{-3mm}   
\begin{center} 
	\rule{10cm}{0.1mm} \ \ $(\forall_{L})$  
	
 $\Delta', \forall x D; \Gamma  \, \vlargo \, 
	\exists w G_{1}$   
\end{center}
\vspace{-3mm}  

where:\\
-- $y$ is not free in $\Delta'$, $\forall x D$, $\Gamma$, $\exists w 
 G_{1}.$\\[-1mm]
-- $z$ is not free in $\Delta'$, $D[y/x]$, $\Gamma$, $C'$, $\exists w 
 G_{1}.$\\[-1mm]
--  $u$ is a new variable. \\

We can transform this into the following proof: \\

\quad $\Delta', D[u/x]; \Gamma, C' \wedge C, u \approx y  \, \vlargo \, 
 	G_{1}[z/w] \qquad  
 	\Gamma, C'\wedge C \, \vdash_{\cal C}\,\exists u (u \approx y)$  
 \vspace{-3mm}  
 \begin{center}	 
 \rule{10cm}{0.1mm} \ \ \ $(\forall_{L})$  
 \end{center}
 \vspace{-3mm}  
 \qquad	 $\Delta', \forall x D; \Gamma,  C' \wedge C  \, \vlargo \, 
 	G_{1}[z/w]$ 
 	
\qquad\qquad \qquad $\downarrow$   Lem. \ref{exist} 
  
  \quad	$\Delta', \forall x D; \Gamma, \exists y (C' \wedge C) \, \vlargo \, 
 	 G_{1}[z/w]\qquad\qquad 
 	 \Gamma \, \vdash_{\cal C}\, \exists z \exists y (C' \wedge C)$  
 \begin{center}	  
 \vspace{-3mm}  
 	\rule{10cm}{0.1mm}  \ \ \  $(\exists_{R})$   
	
 $\Delta', \forall x D; \Gamma  \, \vlargo \, 
 	\exists w G_{1}$   
\end{center}
 
 where:\\
--  $z$ is not free in $\Delta'$, $\forall x D$, $\Gamma$, $\exists w 
 G_{1}$.\\[-1mm]
 --  $u$ is not free in $\Delta'$, $\forall x D$, $\Gamma$, $C' \wedge C$, 
 $G_{1}[z/w]$.
 \end{proof}     
  
 The next main theorem follows now as a straightforward consequence 
 of the Proof Transformation Lemma \ref{trans}. 

\begin{tho} [Uniform Proofs] \label{unif}
Every ${\cal IC}$-provable sequent has a uniform proof.
\end{tho}
 \begin{proof} Given an ${\cal IC}$-provable sequent with a 
 proof of size $l$, the existence of a uniform proof is established 
 reasoning by induction on $l$, using  Lemma \ref{trans}. 
  \end{proof}
  
 \subsection{The calculus ${\cal UC}$}
 	
 Now we know that uniform proofs are complete for ${\cal IC}$, and 
 their goal-oriented format renders them close to the goal solving 
 procedure we are looking for. However, as an intermediate step we 
 will present a second proof system
${\cal UC}$
 for {\em HH}(${\cal C}$), which will enjoy three properties:
  \begin{enumerate}
  \item[a)] ${\cal UC}$ and ${\cal IC}$ have the same provable sequents.
\item[b)] ${\cal UC}$ 
builds only \underline{U}niform proofs, and it is parameterized by 
a given 
 \underline{C}onstraint system.
\item[c)]  $\vdash_{\cal UC}$ replaces the left-introduction rules by a 
backchaining mechanism.
\end{enumerate}

${\cal UC}$-derivations are {\em very close}\/ to our intended 
computations. Therefore, the   ${\cal UC}$ system will be very useful 
for designing a sound and complete goal solving procedure in the next 
section.

Provability in ${\cal UC}$ is defined as follows. 
 $\Delta;\Gamma \vdash_{\cal UC} \, G$ if and only if the sequent 
 $\Delta; \Gamma \,\vlargo \, G$ has a proof using the following 
 rules: 

\begin{itemize}
\item {\em Axiom to deal with 
constraints}: 
$$
\frac{\Gamma \, \vdash_{\cal C} \, C}{\Delta; \Gamma \,\vlargo \, C}\;(C_{R})
$$

\item {\em Backchaining rule for atomic goals}:
$$
\frac{\Delta; \Gamma \,\vlargo \, \exists x_{1}\ldots \exists x_{n}((A 
\approx A')\wedge G)}{\Delta; \Gamma \,\vlargo \, A'}\;(Clause)
$$
 where $A$, $A'$  begin with  the same predicate 
 symbol and $\forall x_{1}\ldots \forall x_{n}(G \Rightarrow  A)$ is 
 a variant of a formula of $elab(\Delta)$, where $x_{1},\ldots, x_{n}$ do not 
 appear free in the sequent of the conclusion.

\item {\em Rules introducing the connectives and quantifiers of the 
goals}:
\begin{center}
 $(\vee_R)$,
  $(\wedge_R)$,
  $(\Rightarrow_R)$,
  $(\Rightarrow  \! C_{R})$,      
  $(\exists_R)$, 
   $(\forall_R)$.
\end{center}
Defined as in the system ${\cal IC}$.

\end{itemize}

The structure of the rule $(Clause)$, that encapsulates a backchaining 
mechanism, corresponds to the method by which atomic goals, $A'$, 
will be solved by the goal solving procedure to be presented in 
Section \ref{Solve}. As usual in logic programming, an ``instance'' of a 
clause with 
head $A$ and body $G$ is searched, in such a way that $A \approx A'$ 
and $G$ can be proved. By the definition of ${\cal UC}$, 
the existential quantification on the right 
hand side of the premise sequent forces a search for this ``instance'' 
(managed by means of constraints in our system). Note that a similar 
behaviour would result from  the application of $(\forall_L)$ if we 
would make use of  ${\cal IC}$. 

The next auxiliary lemma is needed to show that ${\cal UC}$ and ${\cal IC}$ 
have the same deductive power. It can be viewed as a particular kind of cut 
elimination for ${\cal IC}$, where the cut formula is taken from the 
elaboration of 
the program in the left side of the sequent. We cannot apply directly 
any classical cut elimination result, because constraint entailment is 
embedded into our proof system.

\begin{lemm}[Elaboration] \label{elab}
For any $\Delta, \Gamma, A$ and $F \in elab(\Delta)$:
if $\Delta, F;\Gamma \, 
\vdash_{\cal IC}\, A$, 
then $\Delta;\Gamma \, \vdash_{\cal IC}\, A$.
\end{lemm}
\begin{proof} It appears in the Appendix.
\end{proof} 

Now we can prove the promised equivalence between
${\cal UC}$ and ${\cal IC}$.
\begin{tho}
The proof systems ${\cal IC}$ and ${\cal UC}$ are equivalent. That means,
for any program $\Delta$, for any set of constraints
$\Gamma$, and for any goal $G$ it holds:
\begin{center}$\Delta; \Gamma\, \vdash_{\cal IC}\, G$
if and only if
$\Delta; \Gamma\, \vdash_{\cal UC}\, G.$\end{center}
\end{tho}
\begin{proof} We prove both implications by induction on the 
size of proofs.
\begin{enumerate}
\item[$\La$)]
Assuming $\Delta; \Gamma\, \vdash_{\cal IC}\, G$,
we prove  $\Delta; \Gamma\, \vdash_{\cal UC}\, G$ by case analysis on 
the structure of $G$.

     If $G \equiv A$, by the 
	   Proof Transformation Lemma (\ref{trans}) there are 
	  $n$ ($n \geq 0$) constraints $C_{1},\ldots,  C_{n}$,
	  a variant $\forall x_{1}\ldots \forall x_{n}(G'\Rightarrow  A')$ 
	  of some formula of $elab(\Delta)$,   
 with $x_{1},\ldots, x_{n}$ 
 new distinct variables,
 $x_{i}$  not appearing free in $C_{1},\!\ldots ,$ $ C_{i-1}$, for $1<  i 
\leq n$, and 
$A$, $A'$  beginning with  the same predicate 
 symbol,  such that:
	   \begin{enumerate} 
 
  \item[(a)] $\Gamma\,	\vdash_{\cal C}	\, \exists x_{1} C_{1}$;\hspace{2mm}
 	$\Gamma, C_{1}\, \vdash_{\cal C} \,	\exists	x_{2} C_{2};\hspace{2mm}  
 	 \ldots ; \hspace{3mm}	
  \Gamma,\,C_{1},	\ldots,	C_{n-1}\vdash_{\cal	C} \, \exists x_{n}	C_{n}$.

 \item [(b)] $\Gamma, C_{1},\ldots, C_{n}
  \, \vdash_{\cal C} \,A'\approx A$.
 
 \item [(c)] $\Delta;\Gamma, C_{1},\ldots, C_{n}
  \, \vdash_{\cal IC} 
 \, G',$ with a shorter proof, or $G' \equiv \top$.
  
\end{enumerate}
By (b) and $(C_{R})$, $\Delta; \Gamma, 
C_{1},\ldots, C_{n}\, \vdash_{\cal UC} 
 \,A'\approx A$. By (c) and the induction hypothesis,  
$\Delta;\Gamma,C_{1},\ldots, C_{n} \, \vdash_{\cal UC} \, G'$. Note 
that if $G' \equiv \top$, the proof of this sequent is a direct consequence 
of the rule $(C_{R})$. So 
applying $(\wedge_{R})$, 
$\Delta, \Gamma, C_{1},\ldots, C_{n}
 \, \vdash_{\cal UC} 
 \,(A'\approx A)\wedge G'$. Now, in accordance with (a) and the 
 conditions on $x_{1},\ldots, x_{n}$,
  it is possible to apply 
$(\exists_{R})$ $n$ times obtaining
$\Delta; \Gamma \, \vdash_{\cal UC} 
 \,\exists x_{1}\ldots \exists x_{n}((A'\approx A)\wedge G')$. 
 Therefore, using $(Clause)$, $\Delta; \Gamma\, \vdash_{\cal UC}\, A$.\\
 The cases for non atomic formulas are immediate due to the  
  Proof Transformation Lemma (\ref{trans}), the definition of ${\cal UC}$ and the induction 
 hypothesis.

\item[$\Leftarrow$)] Let us also prove only the atomic case, the 
others are 
proved using the induction hypothesis and the definition of the 
calculi ${\cal UC}$, ${\cal IC}$.

Assume $\Delta; \Gamma\, \vdash_{\cal UC}\, A$,
	then by the definition of ${\cal UC}$ the rule $(Clause)$ has been 
	applied and
	$\Delta; \Gamma \,\vdash_{\cal UC} \, \exists x_{1}\ldots \exists 
	x_{n}((A' \approx A)\wedge G'),$ with a shorter proof, where  
$\forall x_{1}\ldots \forall x_{n}(G' \Rightarrow  A')$ is 
 a variant of a formula of $elab(\Delta)$ with $x_{1},\ldots, x_{n}$ 
 new variables and 
$A$, $A'$  beginning with  the same predicate 
 symbol. Because of the form of ${\cal UC}$'s inference rules, the 
 only way to derive this sequent is by $n$
 successive  applications 
 of $(\exists_R)$. Since $x_{1},\ldots, x_{n}$ are new\footnote{Without
  loss of generality we can consider that 
$x_{i}$ does not appear free in $C_{1}, \ldots, C_{i-1}$, for $1<  i 
\leq n$.}, we can assume:
 \begin{enumerate} 
\item[(a)] $\Gamma\, \vdash_{\cal C} \, \exists x_{1} C_{1}$;\hspace{3mm}
 $\Gamma, C_{1}\, \vdash_{\cal C} \, \exists x_{2} C_{2};\hspace{3mm}
  \ldots;  \hspace{3mm}
\Gamma,\,C_{1}, \ldots, C_{n-1}\vdash_{\cal C} \, \exists x_{n} 
C_{n}$.

 \item [(b)] $\Delta;\Gamma, C_{1},\ldots,C_{n} \, \vdash_{\cal UC} 
 \,(A'\approx A) \wedge 
 G',$ with a shorter proof.
  
\end{enumerate}
 Then by (b) and according to the definition of ${\cal UC}$,
 $\Delta;\Gamma, C_{1},\ldots,C_{n} \, \vdash_{\cal UC} 
 A'\!\approx\! A$ and 
 $\Delta;\Gamma, C_{1},\ldots,C_{n} \, \vdash_{\cal UC} 
 G'$ with shorter proofs. Therefore,
 by the induction hypothesis,
 \begin{center}
 $\Delta;\Gamma, C_{1},\ldots,C_{n} \, \vdash_{\cal IC} 
 \,A'\approx A$  ($\dag$) and\\
 $\Delta;\Gamma, C_{1},\ldots,C_{n} \, \vdash_{\cal IC}$ 
 $G'$ ($\ddag$).
 \end{center}
 ($\dag$) implies  
 $\Gamma, C_{1},\ldots,C_{n} \, \vdash_{\cal C} 
 \,A'\approx A$, by the Proof Transformation 
 Lemma (\ref{trans}). Then,  by $(Atom)$,
 $$\Delta, A';\Gamma, C_{1},\ldots,C_{n}
  \, \vdash_{\cal IC} \,A \; \; (\diamond),$$ so applying 
   $(\Rightarrow_{L})$ to ($\ddag$) and ($\diamond$),
  $$\Delta, G'\Rightarrow A';\Gamma, C_{1},\ldots,C_{n}
  \, \vdash_{\cal IC} \,A.$$ 
   Now by $n$ applications of $(\forall_{L})$, using (a) and the 
   conditions on $x_{1} \ldots, x_{n}$, we obtain
  $$\Delta, \forall x_{1}\ldots \forall x_{n}
  (G'\Rightarrow A');\Gamma
  \, \vdash_{\cal IC} \,A.$$
  \end{enumerate}
 \hspace*{8mm} Therefore by the  Elaboration Lemma (\ref{elab}) $\Delta;\Gamma
  \, \vdash_{\cal IC} \,A.$ 
\end{proof}

The properties stated in Lemma \ref{cons} and Lemma \ref{exist} hold 
also for ${\cal UC}$-derivability. This is ensured by the next two 
lemmas that are proved in the Appendix. \\
 
 \begin{lemm}\label{ucons} For any $\Delta, \Gamma,  G$,
if 
$\Gamma'$ is a set of constraints such that
$\Gamma'\, \vdash_{\cal C}\, \Gamma$, and
$\Delta; \Gamma\, \vdash_{\cal UC}\, G$, then 
$\Delta;\Gamma'\, \vlargo\, G$ has a ${\cal UC}$-proof of the same size. 
\end{lemm}

\begin{lemm}\label{uexist}
 For any $\Delta, \Gamma, C, G$,
if $\Delta; \Gamma, C\, \vdash_{\cal UC}\, G$ and $x$ is a variable that 
does not appear free in $\Delta, \Gamma, G$,
then 
$\Delta; \Gamma, \exists x C\, \vlargo\, G$ has a 
${\cal UC}$-proof of the same size. 
\end{lemm}

 From now on we will work only with the calculus ${\cal UC}$.

 \section{ A Goal Solving Procedure} \label{Solve}

We now turn to the view of {\em HH}(${\cal C}$)
as a logic programming language. Solving a goal $G$ using a program 
$\Delta$ means finding
a ${\cal C}$-satisfiable constraint $R$ such that 
$$\Delta; R \vdash_{\cal UC} G.$$ Any constraint $R$ with this 
property is 
called a {\em correct answer constraint}. For instance, $R \equiv x^2 
\leq 1/2$
is a correct answer constraint for the $disc$ example, as shown in the 
introduction.

We will present a goal solving procedure as a transition system. 
Goal solving 
will proceed by transforming an initial state through a sequence of 
intermediate states, ending in a final state. Each state will 
conserve 
the goals that remain to be solved and a partially calculated answer 
constraint. 
The final state will not have any goal to be solved. In the following 
we will 
formalize these ideas and show soundness and completeness of 
the proposed procedure.

\begin{de} A {\em state} w.r.t. a finite set of variables $V$, 
 written  ${\cal S}$, has the form $\Pi [S  \Box {\cal G}]$ where:
 ${\cal G}$ is a multiset of triples $\langle\Delta,C, G\rangle$ 
($\Delta$ {\em local program},
 $C$ {\em local constraint} formula and 
 $G$ {\em local goal}).
$\Pi$ is  a   quantifier  prefix $Q_{1} \! x_{1}\!
        \ldots \! Q_{k}\! x_{k}$ where $x_{1}, \ldots, x_{k}$ are distinct 
variables
        not belonging to  $V$, and every $Q_{i}$, $1 \leq i \leq k$, 
is the quantifier 
        $\forall$ or $\exists$.  
 $S$ is a {\em global constraint} formula.
\end{de} 
This complex notion of state is needed because the goal solving 
transformations, presented below, introduce local clauses and local 
constraints. Of course, local clauses also arise in {\em HH}, see 
\cite{Nad93}. Initial states are quite simple as can be seen in 
Definition \ref{initial}.

We say that a {\em state} $\Pi[S \Box {\cal G}]$ is {\em 
satisfiable} 
iff the associated constraint  
formula  $\Pi S$, also called {\em partially calculated answer 
constraint}, is ${\cal C}$-satisfiable.

If $\Pi'$, $\Pi$ are quantifier prefixes such that 
$\Pi'$ coincides with the first $k$ elements  of $\Pi$,  $0 \leq k 
\leq n$,
where $n$ is the  number of elements of $\Pi$, then  
$\Pi-\Pi'$  represents the result of eliminating $\Pi'$ of 
$\Pi$.  For instance $\forall x \forall y \exists z \forall u \exists 
v - \forall x \forall y \exists z \equiv \forall u \exists v$.

To represent a multiset ${\cal G}$, we will simply write its elements
separated by commas, assuming that repetitions are relevant but 
ordering is  not. In particular, the notation 
${\cal G},\langle \Delta,C,G\rangle$ stands for any multiset which 
includes at least one occurrence of the triple $\langle \Delta,C,G\rangle$.  
 	
\begin{de}[Rules for transformation of states] \label{str}

The transformations permitting to pass from a state ${\cal S}$ w.r.t. a set of 
variables $V$, to another state ${\cal S'}$ w.r.t. $V$,  
written  as ${\cal S} \vded{\cal S'}$,  
are the following: 

\begin{enumerate}
	\item[{\it i)}]  {\em Conjunction}.
	
$\Pi[S \Box {\cal G},\langle\Delta,C,G_1\wedge 
G_2\rangle]$   	
                 $\vded$
                $\Pi[S \Box {\cal G},\langle \Delta,C,G_1\rangle,
                \langle\Delta,C, G_2\rangle]$. 	

	\item[{\it ii)}] {\em Disjunction}.
	
	 $\Pi[S \Box {\cal G},\langle\Delta,C,G_1\vee 
G_2\rangle]$  $\vded$
                $\Pi[S \Box {\cal G},\langle\Delta,C,G_i\rangle]$,
                for $i = 1$ or $2$\\ (don't know choice). 	

	\item[{\it iii)}] {\em Implication with local clause}.

 $\Pi[S \Box {\cal G},\langle\Delta,C,D\Rightarrow 
G\rangle]$   	
                 $\vded$
                $\Pi[S \Box {\cal G},\langle\Delta \cup 
\{D\},C,G\rangle]$. 	

	\item[{\it iv)}] {\em Implication with local constraint}.

 $\Pi[S \Box {\cal G},\langle\Delta,C,C'\Rightarrow 
G\rangle]$   	
                 $\vded$
                $\Pi[S \Box {\cal G},\langle\Delta,C\wedge 
C',G\rangle]$. 	

	\item[{\it v)}] {\em Existential quantification}.
	
	 $\Pi[S \Box {\cal G},\langle\Delta,C,\exists x 
G\rangle]$   	
                 $\vded$
                $\Pi\exists w [S \Box {\cal G},\langle\Delta,C,G[w/x]
                \rangle]$,\\ 
 where $w$ does not appear in $\Pi$ nor in $V$. 	

	\item[{\it vi)}] {\em Universal quantification}.

 $\Pi[S \Box {\cal G},\langle\Delta,C,\forall x  
G\rangle]$   	
                 $\vded$
                $\Pi \forall w [S \Box {\cal G}, 
             \langle\Delta,C,G[w/x]\rangle]$,\\
 where $w$ does not appear in $\Pi$ nor in $V$. 	

	\item[{\it vii)}] {\em Constraint}.
	 
	 $\Pi[S \Box {\cal G},\langle\Delta,C,C'\rangle]$   	
                 $\vded$
                $\Pi[S \wedge (C \Rightarrow C') \Box  {\cal G}]$.\\
If  $\Pi(S \wedge (C \Rightarrow 
                C'))$ is ${\cal C}$-satisfiable.   	
                 
	\item[{\it viii)}] {\em Clause of the  program}.
	
        $\Pi[S \Box {\cal G},\langle\Delta,C,A\rangle]$   	
                 $\vded$
                $\Pi 
                [S \Box {\cal G},\langle\Delta,
             C,\exists x_1\ldots \exists x_n ((A' \approx A) \wedge 
             G)\rangle]$.\\
Provided that  $\forall x_1\ldots\forall x_n (G\Rightarrow 
                A')$ is a variant of some clause in  
$elab(\Delta)$ (don't know choice),   
               $x_1,\ldots,x_n$ do not appear in $\Pi$ nor in $V$, 
and  
                $A', A$ begin with the same predicate symbol.
\end{enumerate}
\end{de}
Note that every transformation  can be applied to an arbitrary 
triple $\langle\Delta, C, G\rangle$ within the state, since ${\cal 
G}$ is viewed as a multiset. Moreover, all choices involved in carrying 
out a sequence of state transformations are {\em don't care}, except 
those explicitly labeled as {\em don't know} in transformations {\em 
ii)}
and {\em viii)} above. One can 
 commit to don't care choices without compromising completeness. 
In other words: at the implementation level, backtracking is needed 
only for don't know choices.

The following definition formalizes the setting needed for goal solving. 

\begin{de}\label{initial}
The  {\em initial state} for a program  $\Delta$ and a goal $G$ 
is a state w.r.t. the set of free variables of $\Delta$ and $G$ 
consisting in
  ${\cal S}_0 \equiv [\top \Box \langle\Delta,\top, G\rangle].$

 A {\em resolution of a goal G from a  program} $\Delta$ is a 
finite 
sequence of states w.r.t. the free variables of $\Delta$ and $G$,
${\cal 
S}_0,\ldots, {\cal S}_n$, such that:
\begin{itemize}
\item  ${\cal S}_0$ is the initial state for 
$\Delta$ and $G$.
\item  ${\cal S}_{i-1} \vded {\cal S}_i$, $1 \leq i \leq n$, 
by means of any of the transformation rules.
\item The {\em final state}  
${\cal S}_n$ has the form  $ \Pi_{n}[ S_{n} \Box \emptyset]$.
\end{itemize}
 The   constraint  $\Pi_{n}S_{n}$ is called the  {\em 
answer constraint} of this resolution. 
\end{de}

\begin{ex}
{\rm
Using $\Delta$, $G$ and $R$ as given in the {\it disc example} 
(see the Introduction) it is possible to build a 
resolution of $G$ from  $\Delta$ with answer constraint $R$ as follows: 

[$\top \Box  \langle  \Delta, \top,
 \forall y (y^2 \leq 1/2 \Rightarrow disc \ (x,y) )\rangle]$
$\vded_{vi)}$\\
$\forall y [\top \Box  \langle \Delta, \top, y^2 \leq 1/2 
\Rightarrow disc
 \ (x,y) \rangle ]$
 $\vded_{iv)}$\\
$\forall y [\top \Box  \langle  \Delta,  y^2 \leq 1/2, disc \ 
(x,y)
 \rangle ]$
$\vded_{viii)}$\\
$\forall y [\top \Box  \langle  \Delta,  y^2 \leq 1/2, \exists u 
\exists v
 ( x \approx u \wedge y \approx v \wedge u^2 + v^2 \leq 1/2) \rangle ]$
$\vded_{vii)}$\\
$\forall y [y^2 \leq 1/2 \Rightarrow \exists u \exists v 
( x \approx u \wedge y \approx v \wedge u^2 + v^2 \leq 1 )\Box 
\emptyset]$\\
 since $\forall
 y (y^2 \leq 1/2 \Rightarrow \exists u \exists v
  ( x \approx u \wedge y \approx v \wedge u^2 + v^2 \leq 1 ))$ is 
  ${\cal R}$-satisfiable.
  
 \noindent So the answer constraint is\\
  $\forall y (y^2 \leq 1/2 \Rightarrow \exists u \exists v 
  ( x \approx u \wedge y \approx v \wedge u^2 + v^2 \leq 1 )) 
  \Requiv$\\ 
$\forall y (y^2 \leq 1/2 \Rightarrow x^2 + y^2 \leq 1) \Requiv 
x^2 \leq 1/2.$ 
\fde
}
\end{ex}

For {\em CLP}\/ programs, the goal transformations {\it ii), iii), 
iv)} and {\it vi)} can 
never be applied. Therefore, the state remains of the  form
$\Pi [S  \Box {\cal G}]$, where $\Pi$ includes only existential 
quantifiers 
and ${\cal G}$ is a multiset of triples  $\langle\Delta,C, 
G\rangle$ such that
$\Delta$ is the global program. For states of this kind, the goal 
transformations {\it i), v), vii)} and {\it viii)}
 specify constrained {\em SLD}\/ resolution, 
as used in {\em CLP}; see e.g. \cite{surv,clprep}. On the other hand, 
traditional {\em 
HH} programs can be emulated in our framework by using 
the Herbrand constraint system ${\cal H}$ and avoiding constraints in 
programs and initial goals. Then  transformation {\it iv)} becomes useless, 
and 
the remaining goal transformations can be viewed as a more
abstract formulation of the goal 
solving procedure from \cite{Nad93}. Transformation {\it viii)} 
introduces equational constraints in intermediate goals, and in 
transformation {\it vii)} the local constraint $C$ is simply $\top$. 
Therefore, 
$\Pi(S \wedge (C \Rightarrow C'))$ is equivalent to $\Pi(S \wedge 
C')$, where 
$S \wedge C'$ can be assumed to be a conjunction of equations. 
Checking ${\cal 
H}$-satisfiability of $\Pi(S \wedge C')$ corresponds to solving a 
unification 
problem under a mixed prefix in \cite{Nad93}.

Admittedly, the labeled unification algorithm presented in  \cite{Nad93}
is closer to an actual implementation, while our description of goal 
solving is more abstract. Note, however, that the goal solving 
transformations are open to efficient implementation techniques. In 
particular, when {\it vii)} adds a new constraint to the global constraint $S$, 
the satisfiability of the new partially calculated answer constraint 
should be checked incrementally, without repeating all the work 
previously done for $\Pi S$. 
Of course, delaying the constraint satisfiability checks until the end 
is neither necessary nor convenient.

\subsection{Soundness}

Soundness of the goal solving procedure means that if $R$ is the 
answer 
constraint of a  resolution  of a goal $G$ from a program $\Delta$, 
then the sequent  $\Delta; R \, \vlargo \, G$ has a ${\cal UC}$-proof.

 The soundness theorem is based on two auxiliary results. The first 
 one ensures that states remain satisfiable along any resolution.

\begin{lemm}\label{sound1}
Let   ${\cal S}_0,\ldots, {\cal S}_n$ be a resolution of a goal $G$ 
from a program $\Delta$, and $V$ the set of free variables of 
$\Delta$ and $G$. Then, for any  $i$, $0 \leq i \leq 
n$, if ${\cal S}_i \equiv \Pi_i [S_{i} \Box {\cal G}_i]$, then
the following properties are satisfied:
\begin{enumerate}
   \item The free variables of the formulas of ${\cal G}_i$, 
 and $S_{i}$ are in $\Pi_{i}$ or in $V$.  
	\item  ${\cal S}_{i}$ is satisfiable.
\end{enumerate}
\end{lemm}
\begin{proof}
The first property is a consequence of the procedure used to build 
the prefix of a state. The initial state satisfies it by definition, 
and when passing from  state ${\cal S}_{i-1}$ to state ${\cal 
 S}_{i}$, $1\leq i \leq n$, if we include new  free variables, these 
will be quantified universally or existentially by $\Pi_{i}$.
  
For the second property, note that $S_{0} \equiv \top$ by definition. 
Moreover, for each transformation step  ${\cal S}_{i-1} \vded  {\cal 
S}_{i}$, one of the three following cases applies:
\begin{itemize}
\item   $S_{i} \equiv 
\! \! \!\!\! \! / \;
S_{i-1}$. Then the transition
must correspond to the transformation {\it vii)} which requires
 ${\cal C}$-satisfiability of $\Pi_{i}(S_{i})$.
 \item  $ S_{i} \equiv S_{i-1}$ and $\Pi_{i} \equiv \Pi_{i-1}$.
 This case is trivial.
 \item  $ S_{i} \equiv S_{i-1}$ and $\Pi_{i} \equiv \Pi_{i-1}Q w$,
 where $Q$ is $\forall$ or $\exists$ and $w$ is a new variable not 
 free in $S_{i-1}$, and not occurring in $\Pi_{i-1}$. Under these 
 conditions,   
$$\Pi_{i}S_{i} \equiv \Pi_{i-1} Q w S_{i-1} \Cequiv \Pi_{i-1}  
S_{i-1},$$
\end{itemize}
\hspace*{9mm}and  ${\cal C}$-satisfiability propagates from $\Pi_{i-1}S_{i-1}$
to $\Pi_{i}S_{i}$.  
\end{proof}
The second auxiliary lemma means that correct answer constraints are 
preserved by any resolution step.

\begin{lemm}\label{sound2}
Assume ${\cal S} \equiv \Pi [S \Box {\cal G}]$ and 
${\cal S}' \equiv \Pi\Pi' [S' \Box {\cal G}']$ are two states w.r.t. a 
set of variables $V$, such that ${\cal S} \vded {\cal S}'$. 
If $R'$ is a constraint with its free variables in 
$\Pi\Pi'$ or in $V$, and such that
$R' \vdash_{\cal C} S'$ and for any 
	 $\langle  \Delta',C', G' \rangle \in {\cal G}'$,  
          $\Delta'; R', C' \, \vdash_{\cal UC}\, G'$, then
          $\Pi'R' \vdash_{\cal C} S$ and for any 
	 $\langle  \Delta,C, G \rangle \in {\cal G}$,  
          $\Delta; \Pi' R', C \, \vdash_{\cal UC}\, G$.
\end{lemm}
\begin{proof}
We
 analyze the different cases, according to the transformation applied. 
We show here the first case, the other cases appear in the Appendix.
 \begin{enumerate}
    
 	\item[i)]  {\em Conjunction}. $\Pi'$ is empty and $S \equiv S'$, so
 	$\Pi'R' \vdash_{\cal C} S$ obviously. On the other hand, let 
 	$\langle  \Delta, C,G \rangle \in {\cal G}$:\\
	If $\langle  \Delta, C,G \rangle \in {\cal G}'$,
 	then 
 	$\Delta; \Pi' R', C\, \vdash_{\cal UC}\, G$  by 
hypothesis, since $\Pi' R' \equiv R'$. \\
 	 If $\langle  \Delta,C, G \rangle \notin  {\cal 
 		G}'$, then  
 $G \equiv G_1 \wedge G_2$ and  $\langle  
 \Delta,C, G_1 \rangle$,  $\langle  \Delta, C,G_2 \rangle  \in
 		 {\cal G}'$.
 		Therefore
 		 $\Delta;  \Pi' R', C\, 
 		\vdash_{\cal UC}\, G_1 \, \, {\rm and} \, \,
 		 \Delta; \Pi' R', C\, \vdash_{\cal UC}\, G_2,$
 		  by 
hypothesis, since $\Pi' R' \equiv R'$,
\end{enumerate}
\vspace*{-2mm}
\hspace*{7.5mm} 	and consequently  $\Delta; \Pi' R', C \, \vdash_{\cal UC}\, G,$
 	by	applying ($\wedge_R$). 
 \end{proof}

\begin{tho} [Soundness] Let $\Delta$ be any program.
 If $G$ is a goal such that there is a resolution ${\cal 
 S}_{0},\ldots,{\cal S}_{n}$ of $G$ from $\Delta$ with 
answer constraint  $R \equiv \Pi_{n}S_{n}$, then $R$ is 
${\cal C}$-satisfiable and
$\Delta;R \, \vdash_{\cal UC}\, G.$
\end{tho}
\begin{proof} The proof is direct from the previous 
lemmas. ${\cal C}$-satisfiability of $R$ is a consequence of item 2 
of Lemma \ref{sound1}. Besides using 
 Lemma \ref{sound2} we can  prove, 
 that for $0 \leq i \leq n$,  
$\Delta; (\Pi_{n}-\Pi_{i}) \, S_{n}, C \, \vdash_{\cal UC}\, G,$ for any
$\langle  \Delta, C, G \rangle \in {\cal G}_{i}$, and 
$(\Pi_{n}-\Pi_{i}) \, S_{n} \,  \vdash_{\cal C}\, S_{i}$. The case $i = 0$
of this result assures the theorem.
Let us prove it by induction	on the construction	of ${\cal 
  S}_0,\ldots,	{\cal S}_n$, but beginning from	the	last state.	The	
  base case is obvious because ${\cal	
  G}_n	= \emptyset$ and $(\Pi_{n}-\Pi_{n}) \, S_{n} \,  \vdash_{\cal C}\, S_{n}$
  holds trivially. For the
  induction step, we suppose the result for ${\cal	S}_{i+1},\ldots, 
  {\cal S}_n$,	and	we prove it	for	 ${\cal	S}_i$. Taking $(\Pi_{n}-\Pi_{i+1})\,S_{n}$ as the 
  constraint $R'$ of  Lemma \ref{sound2},
  the induction hypothesis for 
  $i+1$ indicates that the conditions of Lemma \ref{sound2} are 
  satisfied for ${\cal S}'\equiv {\cal S}_{i+1}$, then this lemma 
  affirms that the result is true for ${\cal S}_i$ as we wanted to 
  prove.  
  \end{proof}

\subsection{Completeness}
Completeness of the goal solving procedure states that  given 
a program $\Delta$, and a goal $G$ such that 
$\Delta; R_{0}\, \vdash_{\cal UC}\, G$ for a ${\cal C}$-satisfiable 
constraint $R_{0}$,  there 
is a resolution of $G$ from $\Delta$ with answer constraint $R$ 
that 
is entailed by $R_{0}$ in the constraint system ${\cal C}$, i.e.
$R_{0} \, \vdash_{\cal C}\,R$. Of course this entailment means that 
the computed answer $R$ is at least as general as the given correct 
answer $R_{0}$.
 
In order to prove this result, we introduce a well-founded ordering 
which measures the complexity of proving that a given constraint is  
a correct answer for a given state. The ordering is based on multisets.

\begin{de}
Let $\Delta$ be a program, $G$ a goal, and $C$, $R$, constraints such 
that  $\Delta; R, C\, \vdash_{\cal UC}\, G$, then we define
$\tau_{R}(\Delta, C,G)$  as the size of the shortest 
${\cal UC}$-proof of the sequent $\Delta; R, C\, \vlargo \, G.$ 

Let ${\cal G}$ be a multiset of triples
 $\langle  \Delta, C, G \rangle$.
 We define ${\cal M}_{{\cal G} R}$ as the multiset of sizes 
$\tau_{R}(\Delta, C, G)$, where the multiplicity of 
$\tau_{R}(\Delta, C, G)$ in  
${\cal M}_{{\cal G} R}$ coincides with  the multiplicity of
 $\langle  \Delta,C, G \rangle$ in ${\cal G}$. 
 \end{de}
   We use the notation $<<$ for the well-founded multiset ordering 
   \cite{DM79} induced  by the 
ordering $<$ over the natural numbers.    

Next, we show that as long as a state  can be transformed, the 
transformation can be chosen to yield a smaller state with respect to $<<$, 
while essentially keeping a given answer constraint $R$.
   
\begin{lemm} \label{complet1}
Let  ${\cal S} \equiv \Pi[S \Box {\cal G}]$ 
be a non-final state w.r.t. a set of variables $V$, and let $R$ be a 
constraint  such 
that  $\Pi R$ is ${\cal C}$-satisfiable and
$R \, \vdash_{\cal C}\,S$.  If   
$\Delta; R, C\, \vdash_{\cal UC}\, G$ for all
 $\langle  \Delta,C, G \rangle \in 
{\cal G}$, 
           then we can find a rule transforming ${\cal S}$ in a state 
${\cal S}' \equiv \Pi'[S' \Box {\cal G}']$ (${\cal S}\vded{\cal S}'$) 
and a constraint $R'$ 
  such that: 
\begin{enumerate}
                     
           \item 
           $\Pi R\, \vdash_{\cal C}\,\Pi' R'$ and $R'\, \vdash_{\cal 
           C}\, S'$. 
          
           \item $\Delta'; R', C'\, \vdash_{\cal UC}\, G'$ 
           for all $\langle  \Delta', C', G' \rangle \in {\cal G}'$.
            Moreover
           ${\cal M}_{{\cal G}' R'} << {\cal 
            M}_{{\cal G}  R}$.                                 
\end{enumerate}
\end{lemm}
\begin{proof}  By induction on the structure of  
$G$, where  $\langle  \Delta, C, G \rangle \in 
{\cal G}$, analyzing  cases.  We show here an illustrative case,
the proof for the other cases appears in the Appendix.

If $G$ has the form $\exists x G_1$, applying the 
 	transformation {\em v)} 
 	we obtain ${\cal S'}$. Let $w$ be the variable used in the 
substitution involved in this transformation. $w$ does not appear in 
$\Pi$, $V$, and we can choose it also not free in $R$.  By hypothesis
 	 $\Delta; R, C\, 
 		\vlargo\, \exists x G_1$ has a proof of size  $l$, 
 		then by the definition of ${\cal UC}$, 
 		there is a constraint formula $C_{1}$ such that 
  		$ \Delta; R, C,C_{1}\, \vlargo\, G_1[w/x]$
 		has a proof of size less than $l$ and 
 		$R, C\, \vdash_{\cal C}\, \exists w C_{1}$.
 	Let  $R' \equiv  R \wedge (C \Rightarrow  C_{1})$. 
 	
 	1. $R\, \vdash_{\cal C}\, \exists w (R \wedge (C \Rightarrow C_{1})),$
	since  $w$ is not free in $R$, $C$,
 	and $R, C\, \vdash_{\cal C}\, \exists w C_{1}$,  	 
therefore $\Pi R\, \vdash_{\cal C}\, \Pi \exists w (R \wedge (C \Rightarrow  
C_{1})) \equiv \Pi' R'$.
        Moreover,  $S' \equiv S$,  $R'\, \vdash_{\cal C}\, R$ and 
$R\, \vdash_{\cal 
           C}\, S$ implies
        $R'\, \vdash_{\cal  C}\, S'$.

	2. Let  $\langle \Delta',C', G' \rangle \in {\cal G'}$.
	If $\langle \Delta', C', G' \rangle \in {\cal G}$, then 
 $\Delta'; R, C'\, \vdash_{\cal UC}\, G'$  by hypothesis,
 	and therefore,  using   
 	  $ R' \, \vdash_{\cal C}\, R$ and Lemma \ref{ucons},
	  $\Delta'; R',C'\, \vdash_{\cal UC}\, G'$ and
 	$\tau_{R'}(\Delta', C', G') \leq \tau_{R}(\Delta', C', 
G')$. \\
 		 If $\langle  \Delta', C',G' \rangle \notin {\cal G}$, 
 		 then $G' \! \equiv \! G_1[w/x]$, $\Delta'\! \equiv \!\Delta$ and 
 		 $C' \!
\equiv \! C$. 
$\Delta; R', C \, \vlargo \, G_1[w/x]$ will also  have
a proof of size less than  $l$,
since $\Delta; \! R,\! C,\! C_{1}  \vlargo$ $G_1[w/x]$
 	has such a proof,  due to $R', C\, \vdash_{\cal C}\, R, C, 
C_{1}$ and Lemma \ref{ucons}. So
 $\Delta'; R', C'\, \vdash_{\cal UC}\, G'$ 
 		for all $\langle  \Delta', C',G' \rangle \in {\cal G'}$,
 	$\tau_{R'}(\Delta',  C', G') < 
                     \tau_{R}(\Delta,  C, G)$,
                    and 
               ${\cal M}_{{\cal G'} R'} << {\cal M}_{{\cal G} R}$.   
\end{proof} 
	
\begin{tho}[Completeness] \label{completeness}
Let $\Delta$ be a program, $G$  a goal and $R_{0}$ a ${\cal 
C}$-satisfiable 
constraint such that 
 $\Delta;R_{0} \, \vdash_{\cal UC} \, G$. Then  there is a 
resolution of  
 $G$ from 
 $\Delta$ with answer constraint  $R$ such that 
$R_{0}  \, \vdash_{\cal C}\, R.$
\end{tho}
\begin{proof}
Using Lemma \ref{complet1}, we can build a sequence
${\cal S}_0 \vded {\cal S}_1 \vded
\ldots\vded {\cal S}_n$
 of state transformations, 
(${\cal S}_{i} \equiv \Pi_{i}[S_{i} \Box {\cal 
G}_{i}]$,  $0 \leq i \leq n$), that is a
a resolution of $G$ from $\Delta$, 
and a sequence of constraints $R_0, \ldots, R_n$
satisfying that for all $i$, $1 \leq i \leq n$:  
 \begin{itemize}
 \item 
 $R_{0} \,\vdash_{\cal C}\,
           \Pi_{i} R_{i}$,        
  \item  
   $R_{i} \,\vdash_{\cal C}\,S_{i}$, 
  \item 
  $\Delta';R_i, C' \, \vdash_{\cal UC} \, G'$, for all
 $\langle\Delta', C', G'\rangle \in {\cal G}_i$. 
 \end{itemize}
We use an inductive construction that is guaranteed to terminate 
thanks to the well-founded ordering $<<$. Let 
${\cal S}_0\equiv [\top \Box \langle\Delta, \top, G\rangle]$ be
the initial state for $\Delta$ and $G$, which we know is not final, 
if we take $R_0$ as the constraint given by the theorem's 
hypothesis,  we obtain 
$R_{0} \,\vdash_{\cal C}\,
           \Pi_{0} R_{0} \mbox{ and } R_{0} \,\vdash_{\cal 
           C}\,S_{0}$,
            since $\Pi_{0}$ is empty  and $S_{0} \equiv  \top$.
             Moreover, by hypothesis,   
 $\Delta;R_0 \, \vdash_{\cal UC} \, G$ is satisfied, and then also  
$\Delta;R_0,\top \, \vdash_{\cal UC} \, G$
 because of $R_0, \top \, \vdash_{\cal C} \, R_{0}$ and Lemma \ref{ucons}.
 
Assume the result true for ${\cal S}_0,\ldots,{\cal S}_i$,
if the state ${\cal S}_i$  is not final, 
  then ${\cal S}_{i}$ and $R_{i}$ fulfill the hypothesis of 
Lemma \ref{complet1}, thus there will be a state ${\cal S}_{i +1}$
 (${\cal S}_i \vded {\cal S}_{i +1}$) and 
a constraint $R_{i + 1}$ 
such that $R_{i+1} \,\vdash_{\cal C}\,S_{i+1}$ and
$\Pi_{i} R_i \, \vdash_{\cal C}\, 
\Pi_{i+1}R_{i+1} \; \;(\dag)$   
Furthermore, for all $\langle\Delta', C',
G'\rangle \in {\cal G}_{i + 1}$,  
$\Delta';R_{i +1}, C' \, \vdash_{\cal UC} \, G'$ 
and $ {\cal M}_{{\cal G}_{i+1} R_{i+1}}\! << {\cal M}_{{\cal G}_i 
R_i}$. 
 Therefore, by the induction hypothesis, $R_{0} \,\vdash_{\cal C}\,
           \Pi_{i} R_{i}$, and with $(\dag)$ we obtain 
   $R_{0} \,\vdash_{\cal C}\,
           \Pi_{i+1} R_{i+1}.$ By successive iteration, as $<<$ is  
well-founded, we must eventually get a final state  ${\cal S}_n$ that will 
in fact satisfy
            $R_{0} \,\vdash_{\cal C}\,
           \Pi_{n} R_{n}$ and $R_{n} \,\vdash_{\cal C}\,S_{n}$ 
           and so $R_{0} \,\vdash_{\cal C}\, 
           \Pi_{n}S_{n}$, where
           $\Pi_{n}S_{n} \equiv R$ is the answer constraint of 
           ${\cal S}_0, \ldots, {\cal S}_n$. 
In this way we conclude  $R_{0} \,\vdash_{\cal C}\, R.$
\end{proof}

For {\em HH}(${\cal H}$) programs such that  constraints appear 
neither in the left-hand side of implications
nor in  initial goals, 
Theorem \ref{completeness} implies an alternative formulation of the 
completeness theorem given in \cite{Nad93} for a goal solving 
procedure for 
first-order {\em HH}. In our opinion, 
using constraints and constraint satisfiability instead of 
substitutions and unification under a mixed prefix,
that requires low level representation details, we gain a more 
abstract presentation. For {\em CLP} programs, Theorem 
\ref{completeness} becomes a 
stronger form of completeness, in comparison to the strong 
completeness theorem 
for success given in \cite{Mah87}, Th. 2 (see also \cite{clprep}, Th. 
4.12). 
There, assuming $\Delta; R \models_{\cal C} G$, the conclusion is 
that 
$R \vdash_{\cal C} \bigvee_{i = 1}^{m} R_{i}$ where $R_{1}, \ldots ,
R_{m}$ are answer constraints computed in $m$ 
different resolutions of $G$ from $\Delta$. Example \ref{maher} below 
 was used in \cite{Mah87} to illustrate the need of 
considering 
disjunctions of computed answers. In fact, there is no single 
computed answer 
$R_{0}$ such that $R \vdash _{{\cal H}} R_{0}$. However, this fact 
doesn't 
contradict Theorem \ref{completeness}, because $\Delta; R \vlargo 
G$ is not 
${\cal UC}$-derivable, as we  will	see immediately.
\begin{ex}\label{maher}
{\rm 
This example is borrowed from \cite{Mah87}. It belongs  to the 
instance {\em HH}(${\cal H}$) given by the 
Herbrand 
constraint system. Consider

$\Delta \equiv \{D_{1}, D_{2}\},$ with
$D_{1} \equiv p(a,b)$, 
$D_{2} \equiv \forall x (x \approx \! \!\!\!\!\! / \ \ a  \Rightarrow 
p(x,b)),$

$G \equiv p(x,y),$

$R \equiv y \approx b.$

\noindent Up to trivial syntactic variants, this is a {\em CLP}(${\cal 
H}$)-program. According to the model theoretic semantics of {\em CLP}(${\cal 
H}$), we get $\Delta; R \models_{\cal H} G$, because either $x \approx a$ or 
$x \approx \! \!\!\!\!\! / \ \ a$ will hold 
in each ${\cal H}$-model of  $\Delta \cup 
\{R\}$. In contrast to this, in ${\cal UC}$ we only can derive
 $\Delta; R \wedge x \approx a\, \vlargo \, G$ (using $D_{1}$) and 
$\Delta; R \wedge x \approx\! \!\!\!\!\! / \ \ a\, \vlargo \, G$ (using $D_{2}$).
And it is  easy	to check that both answers 
      $R \wedge x \approx a$ and 
  $R	\wedge x \approx\! \!\!\!\!\!  / \ \ a$ can be	computed by	the	
    goal  solving transformations.
But we do not obtain 
$\Delta; R \, \vdash_{\cal UC} \, G$.  Since 
$R \,\vdash \!\!\!\!\!/_{\cal H}$ $x \approx a$,
$R \,\vdash \!\!\!\!\!/_{\cal H}$ $x \approx \! \!\!\!\!\! / \ \ a$,
neither $D_{1}$ nor $D_{2}$ can be used to build a ${\cal UC}$-proof.\fde
}
\end{ex}
    	
The example shows a difference between the model-theoretic semantics 
used in {\em CLP}\/ \cite{Mah87} and our proof-theoretical semantics, 
based on provability in the calculus ${\cal UC}$. 
The latter deals with the logical 
symbols in goals and clauses according to the inference rules of 
intuitionistic logic. Therefore ${\cal UC}$-provability turns out to 
be  more constructive than {\em CLP}'s model-theoretic semantics, and 
thus closer to constrained resolution. This is the ultimate reason why 
our completeness Theorem \ref{completeness} involves no disjunction of 
computed answers.

As an illustration of the goal solving procedure, we show next the 
detailed resolution of the second goal from Example \ref{mortage1}. 

\begin{ex}\label{mortage2} 
 Let us recall the program and goal from Example \ref{mortage1}.
As usual in programming practice, we write  program 
 clauses $\forall x_{1} \ldots \forall x_{n} (G \La A)$
in  the form $A\Leftarrow G$\footnote{In fact, we have already 
followed this convention in Section \ref{Exampl}.}.

\begin{tabbing}
$\Delta \equiv$ \{ \= {\it mortgage$(P, T, I, M, B)\Leftarrow$}
                     {\it $0 \leq T \wedge T \leq 3 \ \wedge$} \\ 
		        \`{\it TotalInt} $\approx T*(P*I/1200)$ $\wedge$
                           $ B \approx P + {\it TotalInt} - (T*M)$,\\
                \> {\it mortgage$(P, T, I, M, B)\Leftarrow$}
                   {\it $T > 3$ $\wedge$ QuartInt $\approx 3*(P*I/1200) \wedge$}\\
                 \` {\it mortgage}$(P+{\it QuartInt}-3*M,T-3, I, M, B)$\\
		        \> $\}$\\ 

$G \equiv \forall M  \forall P ( 0.9637 \leq P/(6*M) \leq 0.97 \Rightarrow$\\
           \` $\exists I ( 0 \leq {\it Imin} \leq I \leq {\it Imax} \
            \wedge   mortgage(P, 6, I, M, 0))).$
\end{tabbing}
            
            We present a resolution of $G$ from $\Delta$, 
            using the state transformation rules {\em i)} to {\em viii)}
            from Definition \ref{str}:

\begin{tabbing}           
$ [\top\Box  \langle  \Delta, \top, G \rangle]$\\
 \hspace*{1cm} $\vded_{vi)}$\\
 \noindent$\forall M \forall P [\top \Box  
 \langle \Delta, \top, 0.9637 \leq P/(6*M) \leq 0.97\Rightarrow \exists 
 I (0 \leq {\it Imin} \leq I \leq {\it Imax} \ \wedge$\\
          \` ${\it mortgage}(P, 6, I, M, 0)) \rangle ]$\\
 \hspace*{1cm} $\vded_{iv)}$\\         
   $\forall M \forall P [\top \Box 
   \langle  \Delta, 0.9637 \leq P/(6*M) \leq 0.97,$\\
           \` $\exists I (0 \leq {\it Imin} \leq I 
            \leq {\it Imax} \wedge {\it mortgage}(P, 6, I, M, 0)) \rangle 
            ]$\\			
\hspace*{1cm}$\vded_{v)}$\\
            $\forall M \forall P \exists I [\top \Box$ 
            $ \langle  \Delta, 0.9637 \leq P/(6*M) \leq 0.97,$\\
            \` $0 \leq {\it Imin} \leq I \leq {\it Imax} \wedge {\it 
               mortgage}(P, 6,  I, M, 0) \rangle]$\\
\hspace*{1cm}$\vded_{i),vii)}$\\           
            $\forall M \forall P \exists I[0.9637 
            \leq P/(6*M) \leq 0.97\Rightarrow 0 \leq {\it Imin} \leq I \leq 
            {\it Imax}\Box$ \\
             \` $ \langle  \Delta, 0.9637 \leq P/(6*M) \leq 0.97, 
                {\it mortgage}(P, 6, I, M, 0) \rangle]$\\
\hspace*{1cm} $\vded_{viii)}$\\            
            $\forall M \forall P\exists I [0.9637 \leq P/(6*M) \leq 
            0.97\Rightarrow 0 \leq {\it Imin} \leq I \leq {\it Imax} \Box$ \\
              \`$ \langle  \Delta, 0.9637 \leq P/(6*M) \leq 0.97,$\\
              \` $\underbrace{\exists P' \exists T' \exists I' \exists M' 
                    \exists B' \exists {\it QuartInt} (P \approx P' \wedge 6 \approx T' 
                     \wedge  I \approx I'}$\\ 
              \` $\underbrace{\wedge M \approx M' \wedge 0 
                 \approx B' \wedge T' > 3 \wedge {\it QuartInt}\! \approx \! 
                   3\!*\!(P'\!*\!I'/1200)}$\\
              \` $\underbrace{\wedge {\it mortgage}  
                  (P'\! +\! {\it QuartInt} \!-\!3*M', T'\! -\! 
                   3, I', M', B'))} \rangle ] $\\
Simplifying the underbraced formula in the constraint system
            ${\cal R}$, we obtain:\\   \\     
$\forall M \forall P\exists I [0.9637 \leq P/(6*M) 
            \leq 0.97\Rightarrow 0 \leq {\it Imin} \leq I \leq {\it Imax}  
            \Box$ \\
             \` $ \langle  \Delta, 0.9637 \leq P/(6*M) \leq 0.97,$\\
             \` ${\it mortgage}  (P + 3*(P*I/1200)
                         -3*M, 3, I, M, 0) \rangle]$\\ 
\hspace*{1cm}$\vded_{viii)}$\\ 
            $\forall M \forall P \exists I [0.9637 \leq P/(6*M) \leq 0.97\Rightarrow 
            0 \leq {\it Imin} \leq I \leq {\it Imax}\Box$\\
            \` $ \langle  \Delta, 0.9637 \leq P/(6*M) \leq 0.97,$\\
            \` $\underbrace{\exists P'' \exists T'' \exists I'' 
                   \exists M''\exists	 B'' \exists {\it TotalInt} 
                   (P'' \approx P + 3 * (P * I/1200) - 3 * M}$\\ 
            \` $\underbrace{\wedge \ T'' \approx 3 \wedge I'' 
                   \approx I \wedge M'' \approx M \wedge B'' \approx 0 \wedge 0 \leq 
                   T'' \wedge T'' \leq 3 }$\\
            \` $\underbrace{ \wedge \ {\it TotalInt} \approx 
               T''*(P''*I''/1200)  \wedge B'' \approx P'' + TotalInt - 
                 (T''*M''))} \rangle ]$\\ \\
And simplifying anew the underbraced formula in  ${\cal R}$:\\ \\
  $\forall M \forall P\exists I  [0.9637 \leq P/(6*M) \leq 
       0.97\Rightarrow 0 \leq {\it Imin} \leq I \leq {\it Imax} \Box$\\ 
       \` $ \langle  \Delta, 0.9637 \leq P/(6*M) \leq 0.97,$\\
       \` $0 \approx P + 3*(P*I/1200) - 3*M +$\\
	    \` $ 3*(P + 3*(P*I/1200) - 3*M)*I/1200 - 3*M \rangle]$\\ \\
 Applying now transformation {\em vii)}, we obtain the following answer 
 constraint:\\ \\
$\forall M  \forall P\exists I ((0.9637 \leq P/(6*M) \leq 0.97 
     \Rightarrow 0 \leq {\it Imin} \leq I \leq {\it Imax})) \ \wedge$\\ 
	 \` $(0.9637 \leq P/(6*M) \leq 0.97 \Rightarrow 
	        0 \approx P+ 3*P*I/1200 - 3*M+$\\
     \` $3*(P+3*P*I/1200-3*M)*I/1200-3*M))$\\
\hspace*{1cm}$\vdash\!\dashv_{\cal R}$\\            
$\forall M \forall P\exists I (0.9637 \leq P/(6*M) \leq 0.97 
      \Rightarrow 0 \leq {\it Imin} \leq I \leq {\it Imax} \wedge$ \\
     \` $0 \approx P*( 1 + 3 * \frac{I}{1200} + 3*\frac{I}{1200} + 9*\frac 
        {I^2}{1200^2}) - M*(6 + 9*\frac{I}{1200}))$\\
\hspace*{1cm}$\vdash\!\dashv_{\cal R}$\\       
$\forall M  \forall P\exists I (0.9637 \leq P/(6*M) \leq 0.97
       \Rightarrow 0 \leq {\it Imin} \leq I \leq {\it Imax} \wedge $\\ 
      \`  $0 \approx P*(1 + \frac{I}{200} + \frac{I^2}{400^2}) - M*(6 + 
        3*\frac{I}{400}))$\\
\hspace*{1cm}$\vdash\!\dashv_{\cal R}$\\        
$\forall M  \forall P\exists I (0.9637 \leq P/(6*M) \leq 0.97
          \Rightarrow 0 \leq {\it Imin} \leq I \leq {\it Imax} 
          \wedge$\\  
	      \`$ \frac{P}{6*M} \approx \frac{1+ 
           \frac{I}{800}}{1+\frac{I}{200}+\frac{I^2}{400^2}})\equiv 
           C_1$
\end{tabbing}
		   
\noindent We prove $C_1 \vdash\!\dashv_{\cal R} {\it Imin} 
\approx  8.219559\mbox{ (approx.)}\wedge {\it Imax} \approx 10$. In effect, let 
    $$f(I)=\frac{1+ \frac{I}{800}}{1+\frac{I}{200}+\frac{I^2}{400^2}},$$ 
we observe that $f(I)$ is a strictly decreasing continuous function of $I$ for 
any $I \geq 0$, and also that 
     $$f(I) \approx 0.9637 \mbox{(approx.)} \vdash\!\dashv_{\cal 
      R} I \approx 10, \mbox{ and }$$
      $$f(I) \approx 0.97 \vdash\!\dashv_{\cal R} I \approx 
8.219559\mbox{ (approx.)}.$$
 Then, $C_1$ is true iff for any $M$ and $P$ such that 
 $$P/(6*M) \in [0.97 .. 0.9637 \mbox{ (approx.)}],$$ 
 there exists  $I \in 
[\mbox{\it Imax}..\mbox{\it Imin}]$ such that $f(I)\approx P/(6*M)$ 
            ($f$ strictly decreasing continuous function), and this is true 
iff 
            $I$ has its maximum value for $f(I) \approx 0.9637 \mbox{ 
(approx.)}$ and its minimum 
            for  $f(I) \approx 0.97$, or equivalently ${\it Imax} \approx 
10 
            \wedge {\it Imin} \approx 8.219559 \mbox{ (approx.).}$
            \fde 
            
            \end{ex}

\section{Conclusions and Future Work} \label{Concl}
 
 We have proposed a novel combination of Constraint Logic Programming 
({\em CLP}\/) 
 with first-order Hereditary Harrop Formulas ({\em HH}\/). Our 
framework includes a 
 proof system with the uniform proofs property and a sound and 
complete goal 
 solving procedure. Our results are parametric w.r.t. a given 
constraint 
 system ${\cal C}$,  and they can be related to previously known 
results for 
 {\em CLP}\/ and {\em HH}. Therefore, we can speak of a scheme  whose 
 expressivity sums the advantages of {\em CLP}\/ and {\em HH}. 

As far as we know, our work is the first attempt to combine the full 
expressivity 
of {\em HH}\/ and {\em CLP}. A related, but more limited approach, can 
be found 
in \cite{DG}. This paper presents an amalgamated logic that combines 
the Horn 
fragment of intuitionistic logic with the entailment relation of a 
given 
constraint system, showing the existence of uniform proofs as well as 
soundness and completeness of constrained {\em SLD}\/ resolution w.r.t. 
the proof 
system. The more general case of {\em HH}\/ is not studied. Moreover, 
the 
presentation of constrained {\em SLD}\/ resolution is not fully 
satisfactory, because 
the {\em backchaining}\/ transition rule, see \cite{DG},  
guesses an 
arbitrary instance of a program clause, instead of adding unification 
constraints to the new goal, as done in our state transition rule {\em 
viii)}. 

Several interesting issues remain for future research. Firstly, more 
concrete 
evidence on potential application areas should be found. We are 
currently looking for  {\em CLP}\/  applications where greater 
{\em HH}\/ expressivity may be useful, as well as   
for typical {\em HH}\/ applications that can benefit from  the use of numeric 
and/or symbolic 
constraints. Secondly, tractable fragments of our 
formalism (other 
than {\em CLP}\/ and {\em HH}\/ separately) should be discovered. 
 Otherwise, constraint satisfiability and constraint 
entailment may become intractable or even undecidable.  
Our broad notion of constraint system includes any first-order theory 
based on arbitrary equational axiomatization. Such theories are 
sometimes decidable, see \cite{Com93,CHJ94}, but most often 
restricted fragments must be chosen to ensure decidability. 
 Last but 
not least, our framework should be extended to  higher-order 
{\em HH}\/ as 
used in many  $\lambda$-Prolog applications.\\

\noindent {\bf Acknowledgement} We are grateful to the anonymous referees for 
their constructive criticisms.\\

\appendix
\begin{center}
	{\bf Appendix}
\end{center}
\begin{center}
{\bf \em Proofs of results from Section 4.1}
\end{center}

\noindent{\em Lemma \ref{sust}}\\
  For any $\Delta, \Gamma,  G$, $x$ and $t$,
if $\Delta; \Gamma\, \vdash_{\cal IC}\, G$, then there is a proof of the same 
size of
$\Delta[t/x]; \Gamma[t/x] \, \vlargo\, G[t/x]$.
 \begin{proof} By induction on the size $l$ of the proof of the 
 sequent $\Delta; \Gamma\, \vlargo\, G$. 
 
 If $l =1$, then 
 $(C_{R})$ or $(Atom)$ have been applied. In the first case, $G 
 \equiv C$ for some constraint $C$ and $\Gamma\, \vdash_{\cal 
 C}\, C$. Hence $\Gamma[t/x]\, \vdash_{\cal C}\, C[t/x]$, by 
 the properties of $\vdash_{\cal C}$. Therefore the sequent 
 $\Delta[t/x]; \Gamma[t/x] \, \vlargo\, C[t/x]$ has a proof of  
size 1, by applying $(C_{R})$. In the second case, $G \equiv A$,  
for some predicate formula $A$,
$\Delta = \Delta' \cup \{A'\}$, with $A'$ beginning with the same 
predicate symbol as $A$, and $\Gamma\, \vdash_{\cal 
 C}\, A' \approx A$. Hence 
 $\Gamma[t/x]\, \vdash_{\cal C}\,  (A' \approx A)[t/x]$.
  Therefore, applying $(Atom)$,
 $\Delta'[t/x], A'[t/x]; \Gamma[t/x] \, \vlargo\, A[t/x]$ 
 has a proof of  
size 1, and $\Delta[t/x] = \Delta'[t/x] \cup \{A'[t/x]\}$.

If $l>1$, we distinguish cases in accordance with the last rule 
applied in the deduction of $\Delta; \Gamma\, \vlargo\, G$. Let us
analyze some cases (the omitted ones are similar).
\begin{description}

\item[$(\La \! C_{R})$] In this case $G \equiv  C \La G'$, and
 the last step of the proof has the form:
$$\frac{\Delta;\Gamma, C\, \vlargo \, G'}
{\Delta;\Gamma \, \vlargo \, C \La G'}\,\;(\La  \! C_{R})$$
By the induction hypothesis, 
$\Delta[t/x];\Gamma[t/x], C[t/x]\, \vlargo \, G'[t/x]$ has a proof 
of size $l-1$. Then, applying  $(\La  \! C_{R})$,
 we obtain that $\Delta[t/x];\Gamma[t/x] 
\, \vlargo \, (C \La G')[t/x]$ has a proof 
of size $l$. 

\item[$(\forall_{R})$] In this case $G \equiv \forall z G'$ 
and the last step of the proof has 
the form:
$$\frac{\Delta;\Gamma\, \vlargo \, G'[y/z]}
{\Delta;\Gamma \, \vlargo \,\forall z G'}\,\;(\forall_{R})$$
where $y$ does not appear free in the sequent of the conclusion.
We can assume, without loss of generality, that $z \neq x$ and $z$ does 
not appear in $t$. If  this were not the case, the induction hypothesis 
could be applied another time, in order to rename coincident variables.
Also we can assume  that $y$ is different from
$x$ and  that $y$ does not occur in $t$.
 By the induction hypothesis,
$\Delta[t/x];\Gamma[t/x] \, \vlargo \, G'[t/x][y/z]$ has a 
proof of size $l-1$, because under our hypothesis,
 $G'[y/z][t/x] \equiv G'[t/x][y/z]$. Now, applying  $(\forall_{R})$, 
$\Delta[t/x];\Gamma[t/x] \, \vlargo \, \forall z (G'[t/x])$ has a 
proof of size $l$, but this is the expected result because 
$\forall z (G'[t/x]) \equiv (\forall z G')[t/x]$. 

\item[$(\forall_{L})$] In this case $\Delta = \Delta' \cup 
\{\forall z D\}$.  
As before, we can assume that $z \neq x$ and does not appear in $t$,
and the last step of the proof has 
the form:
$$\frac{\Delta', D[y/z];\Gamma, C\, \vlargo \, G \;\;\;\qquad
\Gamma \, \vdash_{\cal C} \, \exists y C}
{\Delta, \forall z D;\Gamma \, \vlargo \,G}\,\;(\forall_{L})$$
where $y$ does not appear free in the sequent of the conclusion.
We can assume without loss of generality that $y$ is different from
$x$ and  that $y$ does not occur in $t$.
Then, by the induction hypothesis,
$$\Delta'[t/x], D[t/x][y/z];\Gamma[t/x], C[t/x] \, \vlargo \, G[t/x]\;\, 
(\dag)$$ has a 
proof of size $l-1$,
because under our hypothesis, $D[y/z][t/x] \equiv D[t/x][y/z]$.
 Now $\Gamma \, \vdash_{\cal C} \, \exists y C$ implies 
$$\Gamma[t/x] \, \vdash_{\cal C} \, \exists y (C[t/x]) \;\; (\ddag),$$ by the 
properties of $\vdash_{\cal C}$ and the fact that 
$(\exists y C)[t/x] \equiv \exists y (C[t/x])$. Then
applying  $(\forall_{L})$ to $(\dag)$ and $(\ddag)$, 
$\Delta[t/x];\Gamma[t/x] \, \vlargo \,  G[t/x]$ has a 
proof of size $l$,  because 
\end{description}
\hspace*{4mm}$\forall z (D[t/x]) \equiv (\forall z D)[t/x]$ and
$\Delta[t/x] = \Delta'[t/x] \cup \{(\forall z D)[t/x]\}$.
\end{proof}
 
\noindent{\em Lemma \ref{cons}}\\
 For any $\Delta, \Gamma,  G$,
if 
$\Gamma'$ is a set of constraints such that
$\Gamma'\, \vdash_{\cal C}\, \Gamma$, and
$\Delta; \Gamma\, \vdash_{\cal IC}\, G$, then 
$\Delta;\Gamma'\, \vlargo\, G$ has a proof of the same size. 

 \begin{proof}
By induction on the size of the proof of the 
 sequent $\Delta; \Gamma\, \vlargo\, G$,
  by case analysis on the last rule applied, 
 and  using the properties of entailment in  
constraint systems.
It is obvious for proofs of size 1. For proofs of size $l>1$, let us 
analyze the case $(\forall_{L})$ (the others are similar). In this 
case, the last step of the proof is of the form:
$$\frac{\Delta', D[y/x];\Gamma, C\, \vlargo \, G \;\;\; \qquad
 \Gamma \, \vdash_{\cal C} \, \exists y C}
{\Delta', \forall x D;\Gamma \, \vlargo \, G}\,\,(\forall_{L})$$
where $y$ does not appear free in the sequent of the conclusion,
and 
 $\Delta = \Delta' \cup \{\forall x D\}$.
 By the induction hypothesis 
 $$\Delta', D[y/x];\Gamma', C\, \vlargo \, G \;\; (\dag)$$ 
 has a proof of size $l-1$.
 We know that $\Gamma \, \vdash_{\cal C} \, \exists y C$, and by the 
 hypothesis  $\Gamma' \, \vdash_{\cal C} \, \Gamma$, so
 $$\Gamma' \, \vdash_{\cal C} \, \exists y C\;\; (\ddag).$$ 
 We can assume that  $y$ does not appear free in $\Gamma'$, in other 
 case, by Lemma \ref{sust}, we can work with  
 $\Delta', D[y'/x];\Gamma', C[y'/y]\, \vlargo \, G$ ($y'$ new), instead of 
 $(\dag)$, and with
 $\Gamma' \, \vdash_{\cal C} \, \exists y' C[y'/y]$, instead of 
 $(\ddag)$, by the properties of  $\vdash_{\cal C}$.
 Then we finish by
 applying $(\forall_{L})$ to $(\dag)$ and $(\ddag)$.
\end{proof}

\noindent {\em Lemma \ref{exist}}\\
 For any $\Delta, \Gamma, C, G$,
if $\Delta; \Gamma, C\, \vdash_{\cal IC}\, G$ and $x$ is a variable that 
does not appear free in $\Delta, \Gamma, G$,
then 
$\Delta; \Gamma, \exists x C\, \vlargo\, G$ has a proof of the same 
size. 

 \begin{proof}
By induction on the size of the proof. We will assume that $x$ appears 
free in $C$, if not $\exists x C\,\vdash_{\cal C}\, C$, and the proof 
is immediate due to Lemma \ref{cons}.

If $\Delta; \Gamma, C\, \vlargo\, G$ has a proof of size 1,
$(Atom)$ or $(C_{R})$ has been applied. In both cases 
$\Gamma, C\, \vdash_{\cal C}\, C'$ for certain constraint $C'$. Both 
$C'$ and $\Gamma$ do not contain free occurrences of  $x$, hence
$\Gamma, \exists x C\, \vdash_{\cal C}\, C'$, and therefore
$\Delta; \Gamma, \exists x C\, \vlargo\, G$ has a proof of size 1. 
 If $\Delta; \Gamma, C\, \vlargo\, G$ has a proof of size 
 $l> 1$, let us discuss some of the possible cases.
 \begin{description}
 \item[$(\exists_{R})$] Then   
 $G \equiv \exists z G'$ and the last step of the proof is
 of the form:
$$\frac{\Delta;\Gamma, C, C'\, \vlargo \, G'[y/z]\;\;\; \qquad
\Gamma, C \, \vdash_{\cal C} \, \exists y C'}
{\Delta;\Gamma, C \, \vlargo \,\exists z G'}\,\,(\exists_{R})$$
where $y$ does not appear free in the sequent of the conclusion. 
Hence, by 
Lemma \ref{cons}, $\Delta;\Gamma, C \wedge  C'\, \vlargo \, G'[y/z]$ has 
a proof of size $l-1$. Now, the conditions 
on $y$ imply that $x \neq y$, so $x$ is not free in  $G'[y/z]$, 
because it is not free in $\exists z G'$.
Then, by the induction hypothesis and again using Lemma \ref{cons}, 
$$\Delta;\Gamma, \exists x C, \exists x (C \wedge C')\, \vlargo \, G'[y/z] \; 
\; (\dag)$$
has a proof of  size $l-1$.
On the other hand,  $\Gamma, C \, \vdash_{\cal C} \, \exists y C'$ 
implies that 
$\Gamma, C \, \vdash_{\cal C} \, C \wedge \exists y C'$ 
 so 
$\Gamma, \exists x C \, \vdash_{\cal C} \, \exists x(C \wedge \exists y
C')$,
since $x$ is not free in $\Gamma$, thus 
$$\Gamma, \exists x C \, \vdash_{\cal C} \, \exists y\exists x(C \wedge 
C')\;\; (\ddag),$$ since $y$ is not free in $C$. 
Therefore  the desired result is obtained by applying $(\exists_{R})$ to 
($\dag$) and $(\ddag)$.

 \item[$(\forall_{R})$] Then  $G \equiv \forall z G'$, and the last step of the 
proof has the form:
$$\frac{\Delta;\Gamma, C\, \vlargo \, G'[y/z]}
{\Delta;\Gamma, C \, \vlargo \,\forall z G'}\,\;(\forall_{R})$$
where $y$ does not appear free in the sequent of the conclusion. Then
$y$ does not occur free in $C$, so it is different from $x$.
Applying the induction hypothesis to the sequent
$\Delta;\Gamma, C\, \vlargo \, G'[y/z]$, we obtain that  
$\Delta;\Gamma, \exists x C\, \vlargo \, G'[y/z]$ has a proof of 
size $l-1$. Then $\Delta;\Gamma,\exists x C\, \vlargo \, G$ has a 
proof
\end{description}
\vspace*{-2.3mm}
\hspace*{3.5mm}of size $l$ by  $(\forall_{R})$.
\end{proof}

\begin{center}
{\bf \em Proofs of results from Section 4.2}
\end{center}

\noindent{\em  Lemma \ref{trans} (Proof Transformation)}\\
 If $G$ is a goal, 
$\Delta$ a program and  $\Gamma$ a  set of constraint formulas, such that   
$\Delta;\Gamma \, \vlargo\, G$ has a proof of size $l$, then:
\begin{enumerate}
	\item  For $G \equiv A$,
		there are  $n$ constraint formulas $C_{1},\ldots,C_{n}$ ($n \geq 0$)
	and a formula $\forall x_{1}\ldots \forall x_{n}$ $(G' \Rightarrow A')$
	that is a variant of some formula in $elab(\Delta)$ such that
	$x_{1},\ldots\!,\! x_{n}$ are new distinct variables not appearing 
	free in $\Delta,\Gamma, A$, where 
	$x_{i}$ does not appear free in $C_{1}, \ldots \!,\! C_{i-1}$, for $1<  i 
\leq n$, 
	and $A'$ begins with the same 
	predicate symbol as $A$. In addition it holds: 
	\begin{enumerate}
		\item $\Gamma \, \vdash_{\cal C} \,\exists x_{1}
		C_{1};$ \hspace*{2mm}
		$\Gamma, C_{1} \, \vdash_{\cal C} \,\exists x_{2}
		C_{2};\ldots; \hspace*{2mm}
		\Gamma, C_{1},\ldots,C_{n-1} \, \vdash_{\cal C} \,\exists x_{n}
		C_{n}$.
	
		\item 
           $\Gamma, C_{1},\ldots,C_{n} \, \vdash_{\cal C} \, A'  \approx 
           A.$       
         \item $\Delta;\Gamma, C_{1},\ldots,C_{n} \, \vlargo\, 
         G'$
	 has a proof of size less than  $l$, or $G' \equiv \top$.
	
	\end{enumerate}
	
        \item If  $G \equiv C$, then  $\Gamma \, \vdash_{\cal C} \, C$.

	\item  If $G \equiv G_1\wedge G_2$, then  $\Delta;\Gamma \, 
	\vlargo\, G_1$ and $\Delta;\Gamma \, \vlargo\, G_2$ have proofs of 
size less than $l$.
 
	\item  If $G \equiv G_1\vee G_2$, then  $\Delta;\Gamma \, 
	\vlargo\, G_i$ has a proof of size less than $l$ for $i=1$ or $2$.

	\item  If $G \equiv D\Rightarrow G_1$, then $\Delta, D;\Gamma \, 
	\vlargo\, G_1$ has a proof of size less than $l$.  
	
		\item  If $G \equiv C\Rightarrow G_1$, then $\Delta;\Gamma, C \, 
	\vlargo\, G_1$ has a proof of size less than $l$.  

	\item For $G \equiv \exists x G_1$,  if $y$ is a variable not appearing 
	free in $\Delta,\Gamma, G$, then there is a constraint formula
	 $C$ such that:
	 \begin{enumerate}
	 \item $\Gamma \, \vdash_{\cal C} \,\exists y C$.
	 \item  
     $\Delta;\Gamma, C \, 
	\vlargo\, G_1[y/x]$ has a proof of size less than $l$. 
	\end{enumerate} 

	\item  If $G \equiv \forall x G_1$, then 
       $\Delta;\Gamma \, \vlargo\, $ $ G_1[y/x]$ has a proof of size 
less than  $l$, where $y$ is  a variable  that 
does not appear free in $\Delta, \Gamma, G$. 
\end{enumerate}

 \begin{proof} We reason by  
 induction on the size $l$ of a given $\mathcal{IC}$-proof of 
$\Delta;\Gamma \, \vlargo\, G$. 

If $l$ is 1, then $G$ has been proved by a single application of 
axiom $(C_{R})$ or axiom $(Atom)$. In  the former 
case, $G$ is a constraint and item 2 of the lemma holds.
In the latter case  $G$ is  an atomic formula $A$ and
 there is $A' \in \Delta$, 
 beginning with the same predicate symbol that $A$  such 
that $\Gamma \, \vdash_{\cal C}\, A' \approx A$.  But $A' \in 
\Delta$ implies $\top \La A' \in elab(\Delta)$, then conditions 
(a), (b) and (c) of  item 1 are satisfied with $n=0$,  
$G' \equiv \top$.

If $l>1$, let us analyze cases according to the last 
inference rule applied in the 
proof of the sequent $\Delta;\Gamma \, \vlargo\, G$. 
The lemma is obviously true by induction hypothesis
 if the last inference rule introduces on the 
right the main connective or quantifier of the goal. So the problem 
is reduced to the rules $(\wedge_{L}), (\La_{L})$ and
$(\forall_{L})$. 
For each of these three rules, we must analyze  cases according to the
structure of $G$. In each case, it is possible to transform the proof 
by  permuting the application of right and left-introduction  rules, 
in the same way as in \cite{Mil-al}. In our setting, however, the 
treatment of $(\forall_{L})$
gives rise to some new situations.
 We analyze the most interesting cases; the ones
we omit can be treated analogously.
\begin{description}
\item[$(\wedge_{L})$]
 Then we can decompose  $\Delta$ as
$\Delta = \Delta'\cup\{ D_{1} \wedge D_{2}\}$, and the last step of 
the proof is of the form:
$$\frac{\Delta', D_{1}, D_{2};\Gamma\, \vlargo \, G}
{\Delta',D_{1} \wedge D_{2};\Gamma \, \vlargo \, 
G}\,\,(\wedge_{L})$$
\end{description}
\begin{itemize}
\item If $G \equiv G_{1}\vee G_{2}$, then by the induction hypothesis, 
 there is a proof of 
 size less than $l-1$ of 
 $\Delta',D_{1}, D_{2};\Gamma\, \vlargo \,  G_{i}$.
 Applying $(\wedge_{L})$ we obtain a proof of size less or equal 
 $l-1$, so less than $l$, of 
$\Delta',D_{1}\wedge D_{2};\Gamma\, \vlargo \,  G_{i}$
 for $i=1$ or $2$.  
\end{itemize}
\begin{description} 
\item[$(\La_{L})$] Then we can decompose  $\Delta$ as
$\Delta = \Delta'\cup\{ G' \La A\}$, and the last step of 
the proof is of the form:
$$\frac{\Delta';\Gamma\, \vlargo \, G' \; \; \;
\Delta',A;\Gamma\, \vlargo \, G}
{\Delta',G' \La A;\Gamma \, \vlargo \, G}\,\,(\La_{L})$$
\end{description}
\begin{itemize}
\item
If $G \equiv \forall x G_{1}$, then  
$\Delta',A;\Gamma\, \vlargo \, \forall x G_{1}$
 has a  proof of size $l_{1}<l$, and by the induction hypothesis
  there is a proof of 
 size less than $l_{1}$ of $\Delta',A;\Gamma\, \vlargo \,  G_{1}[y/x]$,
 where $y$ is a new variable. Then, using 
 that $\Delta';\Gamma\, \vlargo \, G'$ has a proof of size 
 $l_{2}$,
 $l_{1} + l_{2} = l -1$, and 
 applying 
 $(\La_{L})$, $\Delta',G' \La A;\Gamma\, \vlargo \,  G_{1}[y/x]$
has a proof of size less or equal $l_{1}+l_{2}$ so less than $l$, as we wanted 
to prove.

\item If $G \equiv D \La G_{1}$, then
$\Delta',A;\Gamma\, \vlargo \, D \La G_{1}$
 has a  proof of size $l_{1}<l$, so
  by the induction hypothesis there is a proof of 
 size less than $l_{1}$ of $\Delta',A, D;\Gamma\, \vlargo \,  G_{1}$.
  Then, since $\Delta';\Gamma\, \vlargo \, G'$  has a 
  proof of size $l_{2}$, obviously 
   $\Delta', D;\Gamma\, \vlargo \, G'$ also has a 
  proof of size $l_{2}$, and   $l_{1}+l_{2}<l$. Therefore, using 
 $(\La_{L})$, we obtain that $\Delta',G' \La A, D;\Gamma\, \vlargo \, 
 G_{1}$ has a proof of size  less or equal $l_{1}+l_{2}$, 
 so less than  $l$, as we wanted to prove. 
\end{itemize}
\begin{description} 

\item[$(\forall_{L})$] Then we can decompose  $\Delta$ as
$\Delta = \Delta'\cup\{ \forall x D\}$, and the last step of 
the proof is of the form:
$$\frac{\Delta', D[y/x];\Gamma,C' \, \vlargo \, G\;\;\qquad
\Gamma \, \vdash_{\cal C} \, \exists y C'}
{\Delta', \forall x D;\Gamma \, \vlargo \, G}\,\,(\forall_{L})$$
where  $y$ is not free in the 
sequent of the conclusion, and the sequent $$Q \equiv
\Delta', D[y/x];\Gamma,C' \, \vlargo \, G$$ has a proof of size $l-1$.
\end{description}
\begin{itemize}
\item
If $G \equiv C$, 
then by
 the induction hypothesis applied to $Q$, we know that
 $\Gamma, C' \, \vdash_{\cal C} \, C$. Since  
 $\Gamma \, \vdash_{\cal C} \, \exists y C'$ and $y$ is not free in
 $\Gamma, C$, we conclude that $\Gamma \, \vdash_{\cal C} \, C$, due 
 to the properties of $\vdash_{\cal C}$, that coincides with item 2 of 
 the lemma.
 
\item If $G \equiv C \La G_{1}$, 
then by
 the induction hypothesis applied to $Q$, the sequent 
$${\Delta', D[y/x];\Gamma,C', C \, \vlargo \, G_{1}}$$ has a proof of size
 less than $l-1$. Therefore, since $\Gamma \, \vdash_{\cal C} \, \exists y C'$
  implies
 $\Gamma, C \, \vdash_{\cal C} \, \exists y C'$, and $y$ is not free in 
$C$,
 applying  $(\forall_{L})$, 
	$\Delta', \forall x D;\Gamma, C \, \vlargo \, G_{1}$,  has a proof
of size less or equal than $l-1$ so less than $l$. 	

\item If $G \equiv \exists w G_{1}$, then by applying the induction 
hypothesis to $Q$ we conclude that
 there is $C$ such that
$\Gamma, C'\, \vdash_{\cal C}\, \exists z C$, where $z$ is not free
in $\Delta', D[y/x], \Gamma, C',$ $ \exists w G_{1},$ 
and $$\Delta', D[y/x];\Gamma,C',C \, \vlargo \, G_{1}[z/w] \; \;
 (\dag)$$ 
has a proof of size less than $l-1$.
Since $y$ is not free in $\Delta'$, $G_{1}[z/w]$, applying Corollary 
\ref{ren} to $(\dag)$ we obtain that
 $\Delta', D[u/x];\Gamma,C',C,u \approx y \, \vlargo \, G_{1}[z/w]$,
 where $u$ is a new variable, has a proof of the same size, so 
 by Lemma \ref{cons}, 
 $$\Delta', D[u/x];\Gamma,C' \wedge C, u \approx y \, \vlargo \, G_{1}[z/w]\; \;
 (\ddag)$$
 still with a proof of size less than $l-1$. 
  Now 
 by the properties of the constraint entailment,
 $\Gamma, C' \wedge C \, \vdash_{\cal C} \, \exists u (u \approx y)\; (\S)$. Then,
 since $u$ is not free
 in  $\Delta', \forall x D,\Gamma, C' \wedge C, G_{1}[z/w]$, 
 we apply $(\forall_{L})$ to 
 ($\ddag$) and $(\S)$,
obtaining that $$\Delta', \forall x D;\Gamma,  C' \wedge C \, \vlargo \, 
G_{1}[z/w]$$    
 has a proof of size less than or equal $l-1$. Hence using Lemma 
 \ref{exist} 
 $$\Delta', \forall x D;\Gamma, \exists y ( C' \wedge C) \, \vlargo \, 
 G_{1}[z/w]$$ 
  has a proof of size less than or equal $l-1$, 
because, by the assumptions, $y$ is not free in 
$\Delta', \forall x D, \Gamma, G_{1}[z/w]$.  
 Therefore   we can conclude the 
 result for this case (item 7), taking $\exists y ( C' \wedge C)$ as auxiliary
 constraint. In fact, $\Gamma, C'\, \vdash_{\cal C}\, \exists z C$
 implies $\Gamma, C'\, \vdash_{\cal C}\, \exists z(C' \wedge C)$,
 since $z$ is not free in $\Gamma$, $C'$. Hence 
 $\Gamma, \exists y C'\, \vdash_{\cal C}\, \exists z
 \exists y (C' \wedge C)$, since $y$ is not free in $\Gamma$.
 Finally,  $\Gamma \, \vdash_{\cal C}\, 
 \exists z \exists y (C' \wedge C)$, because
  $\Gamma \, \vdash_{\cal C}\,\exists y C'$.  
\item If $G \equiv A$, then
the induction hypothesis for the sequent  $Q$ assures that
 there are  constraints $C_{1},\ldots,C_{n}$ ($n \geq 0$)
	and a formula $\forall x_{1}\ldots \forall x_{n}(G' \Rightarrow A')$
	that is a variant of a formula in $elab(\Delta'\cup \{D[y/x]\})$,
	where 	$x_{1},\ldots,x_{n}$ are new  variables,
	$x_{i}$  not appearing free in $C_{1}, \ldots, C_{i-1}$, for $1<  i 
\leq n$,  
	$A'$ begins with the same predicate symbol as $A$, and 
	such that:
	\begin{enumerate}
	\item[(i)] $\Gamma,C'  \vdash_{\cal C} \exists x_{1}
		C_{1}$; \hspace*{1.5mm}
		$\Gamma,C', C_{1}  \vdash_{\cal C} \exists x_{2}
		C_{2}; \ldots; \hspace*{1.5mm}
		\Gamma,C', C_{1},\ldots,C_{n-1}  \vdash_{\cal C} \exists x_{n}
		C_{n}$.
		\item[(ii)]
           $\Gamma,C', C_{1},\ldots,C_{n}
            \, \vdash_{\cal C} \, A'  \approx 
           A.$       
         \item[(iii)] $\Delta',D[y/x];\Gamma,
         C', C_{1},\ldots,C_{n} \, \vlargo\, G'$
	 has a proof of size less than  $l-1$, or $G' \equiv \top$. 
	 \end{enumerate}
	 	 In order to establish item 1 of the lemma, we distinguish two 
	 	 cases:
	 \begin{enumerate}	 
	\item[(I)]$\forall x_{1}\ldots \forall x_{n}(G' \Rightarrow A')$
 	 is a variant of a formula in $elab(\Delta')$, or
 	 
 	 \item[(II)]$\forall x_{1}\ldots \forall x_{n}(G' \Rightarrow A')$
 	 is a variant of a formula in $elab(D[y/x])$.
	 \end{enumerate}	  
 \noindent(I). If  $\forall x_{1}\ldots \forall x_{n}(G' \Rightarrow A')$
 	 is a variant of a formula in $elab(\Delta')$, then
 	  $\forall x_{1}\ldots \forall x_{n}(G' \Rightarrow A')$
 	 is a variant of a formula in $elab(\Delta)$. 
	 Taking the following $n$ auxiliary constraints 
      $\exists y (C' \wedge C_{1}), \ldots, \exists y 
      (C' \wedge C_{1}\wedge \ldots  \wedge C_{n})$, we will prove 
	 conditions (a), (b) and (c).
	 \begin{description}
	  	\item[$\bullet$]  For condition (a) we need to prove:
	  
	  $
	  \begin{array}{lc}
	  	\Gamma \, \vdash_{\cal C} \,\exists x_{1}\exists y
		(C' \wedge C_{1}) & (1)  \\
	  	\Gamma,\exists y (C' \wedge C_{1}) \,
		 \vdash_{\cal C} \,\exists x_{2}\exists y (C' \wedge C_{1} \wedge C_{2})
		  & (2)  \\
	  	\vdots & \vdots  \\
	  	\Gamma,\exists y (C' \wedge C_{1}), \ldots,
	  	\exists y (C' \wedge C_{1}\wedge \!\ldots\! \wedge C_{n-1}) 
		 \vdash_{\cal C}\\ 
		 \qquad \qquad \qquad \exists x_{n}\exists y 
		 (C' \wedge C_{1} \wedge\!\ldots\! \wedge C_{n}) & (n)
	  \end{array}$
	  
	  This can be deduced from condition (i) above, as follows:
	  
	 \noindent (1). By (i), $\Gamma,C' \, \vdash_{\cal C} \,\exists x_{1}
		C_{1}$, then 
	  $\Gamma,C' \, \vdash_{\cal C} \,
		C' \wedge \exists x_{1} C_{1}$. Hence  
		 $$\Gamma,\exists y C' \, \vdash_{\cal C} \,\exists y
		(C' \wedge \exists x_{1} C_{1}),$$  since $y$ is not free
		 in $\Gamma$. Therefore
		  $$\Gamma,\exists y C' \, \vdash_{\cal C} \,\exists x_{1}\exists y
		(C' \wedge C_{1}),$$  since $x_{1}$ is not free
		 in $C'$. Now we can conclude (1) because
		 $\Gamma  \, \vdash_{\cal C} \,\exists y C'$.
		 
	\noindent (2). By (i), $\Gamma,C', C_{1} \, \vdash_{\cal C} \,\exists x_{2}
		C_{2}$, then 
	  $\Gamma,C'\wedge C_{1} \, \vdash_{\cal C} \,
		 C' \wedge C_{1}\wedge  \exists x_{2} C_{2}$. Hence 
		 $$\Gamma,\exists y (C'\wedge C_{1}) \, \vdash_{\cal C} \,\exists y
		(C' \wedge C_{1} \wedge  \exists x_{2} C_{2}),$$  since $y$ is not free
		 in $\Gamma$. 
		 Therefore 
		 $$\Gamma,\exists y (C' \wedge C_{1})
		  \, \vdash_{\cal C} \,\exists x_{2}\exists y
		(C' \wedge C_{1} \wedge C_{2}),$$  since $x_{2}$ is not free
		 in $C', C_{1}$. 
		 
		 By a similar reasoning, we can prove $(3)$ to $(n-1)$. \\
	  	
	  \noindent $(n)$. By (i), $\Gamma,C', C_{1},\ldots, C_{n-1}
	  	   \, \vdash_{\cal C} \,\exists x_{n}
		C_{n}$, then 
	  $\Gamma,C'\wedge C_{1} \wedge \ldots \wedge C_{n-1} \, \vdash_{\cal C} \,
		 C' \wedge C_{1}\wedge \ldots \wedge C_{n-1}\wedge \exists x_{n} C_{n}$. 
		 Hence 
		  $$\Gamma,\exists y (C'\wedge C_{1} \wedge \ldots \wedge C_{n-1})
		   \, \vdash_{\cal C} \,\exists y
		(C' \wedge C_{1} \wedge \ldots \wedge C_{n-1}\wedge \exists x_{n} 
		C_{n}),$$
		  since $y$ is not free in $\Gamma$. 
		 Therefore  
		 $$\Gamma,\exists y (C' \wedge C_{1}
		 \wedge \ldots \wedge C_{n-1})
		  \, \vdash_{\cal C} \,\exists x_{n}\exists y
		(C' \wedge C_{1} \wedge \ldots \wedge C_{n-1}\wedge C_{n}),$$
		  since $x_{n}$ is not free
		 in $C', C_{1}, \ldots,C_{n-1}$. Then we deduce  $(n)$ 
		 obviously. 
		 
	  		\item[$\bullet$] For condition (b) we need: 
	  		 $$\Gamma,\exists y (C'\wedge C_{1}),\ldots, 
              \exists y (C' \wedge C_{1}\wedge \ldots\wedge C_{n})
            \, \vdash_{\cal C} \, A'  \approx A.$$ 
            To deduce this from 
	  		 (ii), we note that $y$ is not free in $\Delta'$,
	  		$\Gamma, A$ by assumption. Moreover, $y$
	  		is not free in $A'$, or else it  would be free in $\Delta'$.
	  		 Therefore, (ii)
	  		  implies that 
	  		$$\Gamma, \exists y (C'\wedge  C_{1}\wedge \ldots\wedge C_{n})
            \, \vdash_{\cal C} \, A'  \approx 
           A,$$ which amounts to what we needed.

  	  	\item[$\bullet$] Finally, for condition (c) we assume the interesting case 
  	  	where  $G'$ is not $\top$.
  	  	We need a proof of size less than $l$ for the sequent
  	  	$$\Delta',\forall x D;\Gamma,
  	  	\exists y (C' \wedge C_{1}), \ldots, \exists y (C' \wedge C_{1} \wedge \ldots 
	 \wedge C_{n}) \, \vlargo\, G' \; \; (\dag)$$
  	  	To deduce this, we first choose a fresh variable $u$, and we 
  	  	apply 
  	  	  Corollary \ref{ren} to (iii), thus obtaining that
  	  	 $$\Delta',D[u/x];\Gamma,
         C', C_{1},\ldots,C_{n}, u \approx y \, \vlargo\, G'$$
	 has a proof of size less than  $l-1$.
	 Since 
	 $u$ is new and  
	 $\Gamma, C', C_{1},\ldots,C_{n}  \, \vdash_{\cal C} \,\exists u (u 
	 \approx y)$, we can  apply  $(\forall_{L})$ obtaining that
	 $$\Delta',\forall x D;\Gamma, C', C_{1},\ldots,C_{n} \, \vlargo\, G'$$
	 has a proof of size less  than  $l$.
	 From this, 
	 Lemma  \ref{cons} 
	 and  Lemma \ref{exist} (note that $y$ is not free in 
	 $\Delta',\forall x D, \Gamma, G'$) lead to a proof of size less 
	 than $l$ for
	 $$\Delta',\forall x D;\Gamma, \exists y (C' \wedge C_{1} \wedge \ldots 
	 \wedge C_{n}) \, \vlargo\, G'.$$
	  Another application of Lemma \ref{cons} leads from this to a proof 
	  of size less than $l$ for the sequent $(\dag)$. 
	  \end{description}
	 \noindent (II). If 
	 $\forall x_{1}\ldots \forall x_{n}(G' \Rightarrow A')$
	 is a variant of a formula in $elab(D[y/x])$, then 
	 $\forall y \forall x_{1}\ldots \forall x_{n}(G' \Rightarrow A')$ 
	 is  a variant of a formula in $elab(\forall x D)$, and so it is a 
	 variant of a formula in $elab(\Delta)$. 
	 Then condition (a)  coincides with (i) plus
	   $\Gamma  \, \vdash_{\cal C} \,\exists y C'$, and (b) is equivalent to 
	  (ii). Moreover from (iii) (assuming that $G'$ is not $\top$) we 
	  can deduce that the sequent 
	  $$\Delta',D[u/x];\Gamma,
         C', C_{1},\ldots,C_{n}, u \approx y \, \vlargo\, G'$$ 
        has a proof of size less than $l-1$, because of Corollary
        \ref{ren} ($u$ is chosen as a new variable).
                  Since $\Gamma, C', C_{1},\ldots,C_{n}  \,
           \vdash_{\cal C} \,\exists u (u \approx y)$, we
           can  apply $(\forall_{L})$ and we obtain a proof of size 
           less than $l$ for the sequent
           $$\Delta', \forall x D;\Gamma,
         C', C_{1},\ldots,C_{n} \, \vlargo\, G'.$$   
         \end{itemize}  
\hspace*{8mm}		 That is precisely condition (c).       
        \end{proof}
  
\begin{center}
{\bf \em Proofs of results from Section 4.3}
\end{center}

   \noindent{\em  Lemma \ref{elab} (Elaboration)}\\
For any $\Delta, \Gamma, A$ and $F \in elab(\Delta)$:
if $\Delta, F;\Gamma \, 
\vdash_{\cal IC}\, A$, 
then $\Delta;\Gamma \, \vdash_{\cal IC}\, A$.

\begin{proof}
Since  $F \in elab(\Delta)$, there will be $D \in \Delta$ such that
$F \in elab(D)$. The proof of the lemma is by 
case analysis 
according to the structure of $D$.
\begin{itemize}
	\item  If $D \equiv A'$, then $F \equiv \top \La A'$. We prove
	 $\Delta;\Gamma \, \vdash_{\cal IC}\, A$ by induction 
	 on the size $l$ of the proof of  $\Delta,F;\Gamma \, \vlargo \, A$.
	 If $l = 1$, the proof consists on the application of $(Atom)$, the 
	 form of $F$ implies that it does not take part in this proof. So 
	 there exists $A''\in \Delta$ such that $\Gamma \, \vdash_{\cal C} \, 
	 A'' \approx A$. Therefore  $\Delta;\Gamma \, \vdash_{\cal IC}\, A$, 
	 by $(Atom)$. Assuming now the result for proofs of size less than $l$, 
	 $l >1$, we proceed by case analysis on the last rule applied in the 
	 proof of $\Delta,F;\Gamma \, \vlargo \, A$. Note that it is 
	 only necessary to analyze the left-introduction rules,
	  since the goal is an atomic 
	 formula. For $(\wedge_{L})$ and  $(\forall_{L})$, we note that
	  $F \equiv \top \La A'$ cannot 
	 participate on this step of the proof, instead a formula of 
	 $\Delta$ has been introduced. For instance, for $(\wedge_{L})$,
	  if $D_{1} \wedge 
	 D_{2}$ is the formula introduced, then $\Delta$ is of the form
	 $\Delta' \cup  \{D_{1} \wedge D_{2} \}$, and the last step of the 
	 proof is:
	 $$\frac{\Delta', D_{1}, D_{2},F;\Gamma\,\vlargo \,A}
	 {\Delta', D_{1}\wedge D_{2},F;\Gamma\,\vlargo \,A} \;\;(\wedge_{L}).$$
	 So $\Delta', D_{1}, D_{2},F;\Gamma\,\vlargo \,A$ has a proof of 
	 size less that $l$, and since $F \in elab(\Delta' \cup \{ D_{1}, 
	 D_{2}\})$, $\Delta', D_{1},D_{2};\Gamma \, \vdash_{\cal IC}\, A$, by 
	 induction hypothesis. The result can be obtained now using the rule
       $(\wedge_{L})$.\\
       For the case  $(\La_{L})$, if the introduced formula is $F$ 
       (other cases are proved as before), then the last step of the 
       proof is:
       $$\frac{\Delta;\Gamma\,\vlargo \,\top \; \;\;
       \Delta, A';\Gamma\,\vlargo \, A}
	 {\Delta,F;\Gamma\,\vlargo \,A} \;\;(\La_{L}).$$
	 Since  $A' 
	 \equiv D$ and $D \in \Delta$,    
	the sequent $\Delta, A';\Gamma\,\vlargo \, A$ can be also written as
	  $\Delta;\Gamma\,\vlargo \, A$, and we are done. 

	\item  If $D \equiv D_{1} \wedge D_{2}$, then $F \in elab(D_{i})$ 
	for $i = 1$ or 2. 
	$\Delta, F;\Gamma \, \vdash_{\cal IC}\, A$,  by hypothesis,
	 then it is easy to prove 
	that also 
	$\Delta, D_{1}, D_{2}, F;\Gamma \, \vdash_{\cal IC}\, A$.
	 Hence, applying structural 
	induction hypothesis to $D_{i}$, $\Delta, D_{1}, D_{2};\Gamma \, \vdash_{\cal IC}\, 
	A$. Therefore 
	$\Delta, D_{1} \wedge D_{2};\Gamma \, \vdash_{\cal IC}\, A$,
	in accordance with the rule $(\wedge_{L})$. This is 
	 equivalent to $\Delta;\Gamma \, \vdash_{\cal IC}\, A$,
	 since $D \equiv D_{1} \wedge D_{2}$ and $D \in \Delta$.     
    
	\item If $D \equiv G_{1} \La D_{1}$, then $F \equiv D$, so $F \in 
	\Delta$ and we have $\Delta;\Gamma \, \vdash_{\cal IC}\, A$ directly.

	\item  If $D \equiv \forall x D_{1}$, then $F \equiv \forall x F_{1}$
	and $F_{1} \in elab(D_{1})$. We proceed by induction on the size $l$ 
	of the proof of $\Delta,F;\Gamma \, \vlargo \, A$. The case $l = 1$ 
	is trivial because $F$ cannot take part in the proof. Similarly we 
	can reason the inductive step for the cases 
	$(\wedge_{L})$ and $(\La_{L})$.The interesting case occurs when $(\forall_{L})$ was the last rule 
	applied and $F$ was the introduced formula.
	In this case, the last proof step is of the form:
	$$\frac{\Delta, F_{1}[y/x];\Gamma, C\,\vlargo \, A \;\; \qquad
	\Gamma\,\vdash_{\cal C} \, \exists y C}
	 {\Delta,F;\Gamma\,\vlargo \,A} \;\;(\forall_{L})
	 ,$$
	 where $y$ is not free in the sequent of the conclusion.\\
	  $\Delta, D_{1}[y/x], F_{1}[y/x];\Gamma, C\vdash_{\cal IC}  A$ can be 
	 deduced from $\Delta, F_{1}[y/x];\Gamma,C\vdash_{\cal IC}  A$.
 Then $\Delta, D_{1}[y/x];\Gamma,  C\,\vdash_{\cal IC} \, A$, 
     since the lemma holds for $D_{1}[y/x]$ --simpler than $D$--
     and $F_{1}[y/x] \in elab(D_{1}[y/x])$. 
	 Therefore 
	 $\Delta, \forall x D_{1};\Gamma\,\vdash_{\cal IC} \, A$, by $(\forall_{L})$, 
	 using the fact that $y$ is not free in 
	 $\Delta,\forall x D_{1}, \Gamma, A$, and that
	 $\Gamma\vdash_{\cal C}  \exists y C$.  We con-
	 \end{itemize} 
	 \vspace*{-1.5mm}
	 \hspace*{8mm}
	clude because 
	  $D \equiv \forall x D_{1}$ and $D \in \Delta$.  
	  \end{proof}

 \noindent{\em Lemma \ref{ucons}}\\ 
 For any $\Delta, \Gamma,  G$,
if 
$\Gamma'$ is a set of constraints such that
$\Gamma'\, \vdash_{\cal C}\, \Gamma$, and
$\Delta; \Gamma\, \vdash_{\cal UC}\, G$, then 
$\Delta;\Gamma'\, \vlargo\, G$ has a ${\cal UC}$-proof of the same size. 

 \begin{proof}
By induction on the size of the proof of the 
 sequent $\Delta; \Gamma\, \vlargo\, G$,
  by case analysis on the last rule applied. Using the definition of 
   the system ${\cal 
  UC}$ and  Lemma \ref{cons},  the only interesting case is
  when the last step corresponds to  rule $(Clause)$.  But the proof 
 in this case is a direct consequence of the induction hypothesis. 
 \end{proof}

\noindent{\em  Lemma \ref{uexist}}\\
 For any $\Delta, \Gamma, C, G$,
if $\Delta; \Gamma, C\, \vdash_{\cal UC}\, G$ and $x$ is a variable that 
does not appear free in $\Delta, \Gamma, G$,
then 
$\Delta; \Gamma, \exists x C\, \vlargo\, G$ has a 
${\cal UC}$-proof of the same size. 

\begin{proof} As in the previous lemma, and due now to 
 Lemma \ref{exist}, we can focus the proof on the case $(Clause)$. In this case 
 $G \equiv A$ and the last step of the proof is of the form:
$$\frac{\Delta; \Gamma, C \,\vlargo \, \exists x_{1}\ldots \exists x_{n}((A' 
\approx A)\wedge G')}{\Delta; \Gamma, C \,\vlargo \, A}\;(Clause)$$
where  $A$, $A'$  begin with  the same predicate 
 symbol, and $\forall x_{1}\ldots \forall x_{n}(G' \Rightarrow  A')$ is 
 a variant of a formula of $elab(\Delta)$,  $x_{1},\ldots, x_{n}$ do not 
 appear free in the sequent of the conclusion.

Since $x$ is not free in $\Delta, A$, and 
$\forall x_{1}\ldots \forall x_{n}(G' \Rightarrow  A')$ is 
 a variant of a formula of $elab(\Delta)$, then $x$ is not free in 
$\exists x_{1}\ldots \exists x_{n}((A' 
\approx A)\wedge G')$. Note also, that $x$ is not free in $\Gamma$,
$\Delta$, by assumption, so applying 
the induction hypothesis to the sequent 
$\Delta; \Gamma, C \,\vlargo \, \exists x_{1}\ldots \exists x_{n}((A' 
\approx A)\wedge G')$,  
$$\Delta; \Gamma, \exists x C \,\vlargo \, \exists x_{1}\ldots \exists x_{n}((A' 
\approx A)\wedge G')$$
has a proof of the same size. Hence, applying $(Clause)$, 
$\Delta; \Gamma, \exists x C\, \vlargo\, A$ has a 
${\cal UC}$-proof of the same size that
$\Delta; \Gamma, C \,\vlargo \, A$. 
\end{proof} 

\begin{center}
{\bf \em Proofs of results from Section 5.1}
\end{center}

\noindent{\em Lemma \ref{sound2}}\\
Assume ${\cal S} \equiv \Pi [S \Box {\cal G}]$ and 
${\cal S}' \equiv \Pi\Pi' [S' \Box {\cal G}']$ are two states w.r.t. a 
set of variables $V$, such that ${\cal S} \vded {\cal S}'$. 
If $R'$ is a constraint with its free variables in 
$\Pi\Pi'$ or in $V$, and such that
$R' \vdash_{\cal C} S'$ and for any 
	 $\langle  \Delta',C', G' \rangle \in {\cal G}'$,  
          $\Delta'; R', C' \, \vdash_{\cal UC}\, G'$, then
          $\Pi'R' \vdash_{\cal C} S$ and for any 
	 $\langle  \Delta,C, G \rangle \in {\cal G}$,  
          $\Delta; \Pi' R', C \, \vdash_{\cal UC}\, G$.

\begin{proof}
We
 analyze the different cases, according to the transformation applied. 

 \begin{enumerate}
    
 	\item [{\it ii)}]  {\em Disjunction}.  $\Pi'$ is empty and $S \equiv S'$
 	  as above. Then let us check only the case
 	   $\langle  \Delta, C, G \rangle \notin {\cal G}'$. This implies 
  $G \equiv G_1 \vee G_2$ and  $\langle \Delta,
 	 C, G_1 \rangle \in  {\cal G}'$ or
 		 $\langle  \Delta, C, G_2 \rangle  \in {\cal G}'$.
 		By  hypothesis 		 
$$\Delta; \Pi' R', C \, 
                 \vdash_{\cal UC}\, G_1 \, \, {\rm or} \, \, 
 		\Delta; \Pi' R', C\, \vdash_{\cal UC}\, G_2,$$
 	     since  $\Pi' R' \equiv R'$. Then
 	      $\Delta; \Pi' R', C \, \vdash_{\cal UC}\, G$, 
 		because of the rule ($\vee_R$).

 	\item [{\it iii)}] {\em Implication with local clause}.
 	As before the prefix and the partially calculated answer 
 	constraint do not change. 
 	If $\langle  \Delta, C, G \rangle \notin {\cal G}'$, then    	
 	$G\equiv D \Rightarrow G_1$ and  $\langle  \Delta
 		 \cup \{D\},C, G_1\rangle \in  {\cal G}'$.
 		 Hence, by hypothesis since   $\Pi' R' \equiv R'$, it holds 
 		  $$\Delta, D;\Pi' R', C \, 
 		\vdash_{\cal UC}\, G_1$$
from which we conclude the result by applying ($\Rightarrow_R$).  
 	            
 	\item [{\it iv)}] {\em Implication with local constraint}. 
 	 As in the preceding cases  where there are no changes in $S$ and 
 	 $\Pi$, we check what happens if 
 	 $\langle  \Delta, C, G \rangle \in {\cal G}\setminus{\cal G}'$. In this 
 	 case    	
 	$G \equiv C' \Rightarrow G_1$ and  $\langle  \Delta,
 		C\wedge C', G_1\rangle \in  {\cal G}'$.
 		 By  hypothesis, since 
 		  $\Pi' R' \equiv R'$, we have
 		   $\Delta;\Pi' R', C\wedge C' \, 
 		\vdash_{\cal UC}\, G_1$ then in accordance with Lemma \ref{ucons} 
 		  $$\Delta;\Pi' R', C, C' \, 
 		\vdash_{\cal UC}\, G_1.$$
 	           Now we conclude $\Delta;\Pi' R', C
 	            \, \vdash_{\cal UC}\, G$, by applying ($\Rightarrow  \! C_{R}$).  
 	             	            
  	\item [{\it v)}] {\em Existential quantification}. $\Pi' \equiv \exists 
 	w$ with $w$  a new variable not in $\Pi$ nor in $V$.
 	 Hence, by item {\em i)} of Lemma \ref{sound1},
 	$w$ is not free in the formulas of ${\cal G}$, nor in
 	$S$. Therefore, using the facts $R' \vdash_{\cal C} S'$
 	and $S \equiv S'$, we can conclude $\exists w R' \vdash_{\cal C} S$.
 	 
 	Now let  $\langle  \Delta, C,G \rangle \in {\cal G}$,
	if $\langle  \Delta, C,G \rangle \in {\cal G}'$,
 	 then 
 	$\Delta; R', C\, \vdash_{\cal UC}\, G$,  by 
hypothesis. Then $\Delta; \exists w R', C\, \vdash_{\cal UC}\, G$ 
 	by Lemma \ref{uexist}, because $w$ 
 	is not free in $\Delta, C,G$. \\
       If $\langle  \Delta, C, G \rangle \notin {\cal G}'$,
       $G \equiv \exists x G_1$ and  $\langle  \Delta, 
 	 C, G_1[w/x] \rangle \in  {\cal G}'$.
 		By  hypothesis, $$\Delta; R', C \, 
 		\vdash_{\cal UC}\, G_1[w/x]$$
 		and so also $\Delta; \exists w  R', R', C \, 
 		\vdash_{\cal UC}\, G_1[w/x]$, by Lemma \ref{ucons}.
 		Consequently, applying the rule ($\exists_R$),
 		 $$\Delta;\exists w  R', C \, \vdash_{\cal UC}\, G$$
 		 since 
 		$\exists w  R', C \, \vdash_{\cal C}\, \exists w  R'$,
 		 and $w$ is new for the 
 		sequent of the conclusion.
    
 	\item [{\it vi)}] {\em Universal quantification}. $\Pi' \equiv  \forall w$
 	 with $w$  a new variable w.r.t.  $\Pi$ and $V$, and $S \equiv S'$.
 	 So $\forall w R' \vdash_{\cal C} S$ holds directly
 	 from $R' \vdash_{\cal C} S'$.
 	 
 	Let $\langle  \Delta, C, G \rangle \in {\cal G}$,
	if $\langle  \Delta, C, G \rangle \in {\cal G}'$,
 	 then 
 	$\Delta; R', C\, \vdash_{\cal UC}\, G$,  by  
hypothesis. Then $\Delta; \Pi' R', C\, \vdash_{\cal UC}\, G$ 
 	because 
 	$\Pi' R' \, \vdash_{\cal C}\, R'$ and Lemma \ref{ucons}.\\
 	If $\langle  \Delta, C, G \rangle \notin {\cal G}'$,
 	$G \equiv \forall x G_1$ and 
 		$\langle \Delta,
 		 C, G_1[w/x]\rangle \in  {\cal G}'$. 
 		 By the  hypothesis, since
 		 $\forall w R' \, \vdash_{\cal C}\, R'$
 		 and Lemma \ref{ucons}, we have
		$$\Delta; \forall w R', C \, 
 		\vdash_{\cal UC}\, G_1[w/x]$$   
 		Now, since 
 		  $w$  is not in  $\Pi$ nor in $V$, by item {\em i)} 
 		  of Lemma \ref{sound1},
 		   it is not free in 
$\Delta$, $C$,   $G$, and obviously  $w$ 
 		  is neither free in $\forall w R'$. Then we conclude		     
                      $$\Delta;\forall w R', C\, \vdash_{\cal UC}\, G$$ 
 	 		by applying ($\forall_R$). 
 	 		
 	\item [{\it vii)}] {\em Constraint}. In this case $\Pi'$ is empty  and 
 	$\Pi S' \equiv \Pi (S \wedge (C \La C'))$ is ${\cal C}$-satisfiable.
 	Trivially $R' \vdash_{\cal C} S'$ implies $\Pi' R' \vdash_{\cal C} S$. 

 Now let  $\langle  \Delta, C, G \rangle \in {\cal G}$, the case
  $\langle  \Delta, C, G \rangle \in {\cal G}'$ is easily proved. If
   $\langle  \Delta, C, G \rangle \notin {\cal G}'$, then  	
 	$G \equiv C'$.  $R' \, \vdash_{{\cal C}} \, C \Rightarrow C'$
 	because $R' \vdash_{\cal C} S'$ and $S' \equiv S \wedge (C \La C')$.
 	By the properties of the constraint entailment, we deduce
 	$R', C \vdash_{\cal C} C'$. Then applying the rule ($C_{R}$),
 		$$\Delta;\Pi' R', C \, \vdash_{\cal UC}\, G,$$
 		because $\Pi' R' \equiv R'$.    	
 		 
 	\item [{\it viii)}] {\em Clause of the program}.
                      Since $\Pi'$ is empty and $S \equiv S'$,
                      we only check the case
       $\langle  \Delta, C, G \rangle \in {\cal G}$ and
 	  $\langle  \Delta, C, G \rangle \notin {\cal G}'$. 
 	  In such case  $G \equiv  A$ and there is
 		 $\forall x_1\ldots\forall x_n (G_{1} \Rightarrow A') $ 
 		a variant of  a formula of  $elab(\Delta)$  where:
 			\begin{itemize}  
 		\item
 	 $x_{1},\ldots,\ x_{n}$ 
                      are new variables  not occurring in $\Pi$, 
                      $V$, and therefore not free in  $A$, $\Delta$, 
                      $C$ and $\Pi' R'$. 
 	\item  $A$ and $A'$ begin with the same 
predicate symbol. 
 	 
         \item $\langle  \Delta, C, 
         \exists x_{1}\ldots \exists x_{n}((A' \approx A) \wedge G_{1})
          \rangle \in {\cal G}'$.
          \end{itemize}
By  hypothesis, since  $\Pi' R' \equiv R'$,
	 $$\Delta;\Pi' R', C\,\vdash_{\cal UC}\,
	 \exists x_{1}\ldots \exists x_{n}((A' \approx A) \wedge G_{1}).$$
	 \end{enumerate}
 \hspace*{8mm} Using now the rule $(Clause)$,  we conclude $\Delta; 
\Pi' R', C \, \vdash_{\cal UC}\, G$. 
    \end{proof}

\begin{center}
{\bf \em Proofs of results from Section 5.2}
\end{center} 

\noindent{\em Lemma \ref{complet1}}\\
 Let  ${\cal S} \equiv \Pi[S \Box {\cal G}]$ 
be a non-final state w.r.t. a set of variables $V$, and let $R$ be a 
constraint  such 
that  $\Pi R$ is ${\cal C}$-satisfiable and
$R \, \vdash_{\cal C}\,S$.  If   
$\Delta; R, C\, \vdash_{\cal UC}\, G$ for all
 $\langle  \Delta,C, G \rangle \in 
{\cal G}$,  
           then we can find a rule transforming ${\cal S}$ in a state 
${\cal S}' \equiv \Pi'[S' \Box {\cal G}']$ (${\cal S}\vded{\cal S}'$) 
and a constraint $R'$ 
  such that: 
\begin{enumerate}
                     
           \item 
           $\Pi R\, \vdash_{\cal C}\,\Pi' R'$ and $R'\, \vdash_{\cal 
           C}\, S'$. 
          
           \item $\Delta'; R', C'\, \vdash_{\cal UC}\, G'$ 
           for all $\langle  \Delta', C', G' \rangle \in {\cal G}'$.
            Moreover
           ${\cal M}_{{\cal G}' R'} << {\cal 
            M}_{{\cal G}  R}$.                      
\end{enumerate}

\begin{proof}
 Let us choose any $\langle  \Delta, C, G \rangle \in 
{\cal G}$;  we reason by induction on the structure of  
$G$,  analyzing  cases:
 \begin{itemize}
 
	   \item If	$ G$ has the form  $G_1	\wedge G_2$, $G_1 \vee G_2$,  
 	   $D\Rightarrow G_1$ or	$C_{1}\Rightarrow G_1$,	then we	apply 
  respectively	the	transformation rules {\em i), ii), iii)} or {\em 
  iv)} to	
${\cal  S}$.	Let	${\cal S'}$	be the state obtained after	the	
transformation,	
  and let $R' \equiv  R$:
	   
 	   1. $\Pi R\, \vdash_{\cal	C}\,\Pi' R'$ and $R'\, \vdash_{\cal	
 			  C}\, S'$	are	obvious	by the hypothesis and 
  because $\Pi' \equiv \Pi$,	$S' \equiv S$ and $R' \equiv  R$. 
 			   
 	   2. Let  $\langle	\Delta', C',	G' \rangle \in {\cal G'}$. 
 	   If	$\langle \Delta',	C',	G' \rangle \in {\cal G}$, then	
 		$\Delta';R', C'\,	\vdash_{\cal UC}\, G'$	trivially since	
 	   $\Delta';R, C'\, \vdash_{\cal UC}\, G'$  by hypothesis, 
 		and	$ R' \equiv	R$.	Moreover $\tau_{R'}(\Delta',	C',	G') \!=
 		\tau_{R}(\Delta',	 C', G')$.\\   
 		   If $\langle	\Delta', C',G' \rangle \notin	 {\cal 
 G}$ and {\em i)},	for	example, was applied, then $\Delta' \equiv \Delta$,	
 		$C' \equiv C$, $G \equiv G_1 \wedge G_2$ and $	G'\equiv G_1$ or 
 		$G' \equiv G_2$.   
 		   By hypothesis
 			$\Delta; R,C\, 
 		   \vlargo\, G$	with a proof of	size $l$, therefore	by the 
 		   definition of  ${\cal UC}$, since $R' \equiv R$,
 		   $ \Delta; R', C\, \vlargo \, G_1$ and 
 		   $\Delta; R', C\, \vlargo\, G_2$
 		   have	proofs of size less	than  $l$. Consequently
 		   $\tau_{R'}(\Delta',  C', G_1) < \tau_{R}(\Delta,	C, G)$ and
 		   $\tau_{R'}(\Delta', C', G_2) <	\tau_{R}(\Delta, C, G)$, 
 		   so, finally $\Delta'; R',C'\, \vdash_{\cal	UC}\, G'$ 
 		   for all $\langle	 \Delta',	C',	G' \rangle \in {\cal G'}$ and 
 			   ${\cal M}_{{\cal	G'}	R'}	<< {\cal M}_{{\cal G} R}$. 			   
 			   The argument for transformations {\em ii), iii)} and {\em iv)}
 			   is similar. Note that, in the case of {\em ii)}, we must choose 
 			   $G_{1}$ (resp. $G_{2}$) if the shortest ${\cal UC}$-proof of 
 	$\Delta; R,C\, \vlargo\, G_{1}\vee G_{2}$ contains a subproof of
 	$\Delta; R,C\, \vlargo\, G_{1}$ (resp. $G_{2}$).

 		  \item	 If	 $G$ has the form $\forall x G_1$, we apply	then the 
 	 transformation	rule {\em vi)} and obtain	${\cal S'}$. Assume	$R'	\equiv 	R$:
 	   
 	   1. Trivial since the choice of $w$ assures that
 	   	$\Pi R \Cequiv \Pi \forall w R \equiv \Pi' R'$;
 	   	moreover,		$S' \equiv S$.
 		
 		   2.  Let	 $\langle \Delta', C', G'	\rangle	\in	{\cal G'}$,
 	   if  $\langle	\Delta', C', G' \rangle \in {\cal	G}$, then we  
  obtain		$\Delta';	R',	C'\, \vdash_{\cal UC}\,	G'$, being 
 		$\tau_{R'}(\Delta', C', G') =	\tau_{R}(\Delta',	 C', G')$.\\
 			  If $\langle  \Delta', C', G' \rangle \notin	 {\cal 
 		   G'}$, this is the triple	coming from	the	transformation of 
  $\langle	 \Delta, 
 		   C, G	\rangle$, so   $G' \equiv G_1[w/x]$, $C' \equiv C$
 		    and  $\Delta' \equiv \Delta$. By	hypothesis
 				$\Delta; R, C 
 		   \vlargo\, G$	has	a proof	of size	$l$, then since	$w$	
 		   does	not	appear free	in $\Delta, C, R'	(\equiv R), G_1$,
 		   	because	of  the form of the calculus	${\cal UC}$,
 		   $\Delta; R', C\, 
 		   \vlargo\, G_1[w/x]$	has	a proof	of size	less than $l$, and 
 for that reason $\tau_{R'}(\Delta',  C', G') <	\tau_{R}(\Delta,	 
C,  G)$, and thus we conclude that 2. is valid.
 		   
  	\item  If  $G$ is a constraint $C_{1}$, we apply the 
transformation {\em vii)} obtaining  ${\cal S'}$. Assume $R' \equiv R$:
 	
 	1. $\Pi R\, \vdash_{\cal C}\, \Pi' R'$ is trivial since $\Pi' 
\equiv 
\Pi$. 
 	Furthermore,  
 	$\Delta; R, C \vdash_{\cal UC} C_{1}$ by hypothesis, 
 	so by the definition of  ${\cal UC}$,
 	 $R, C \vdash_{\cal C} C_{1}$ and therefore
 	 $R \vdash_{\cal C} C \Rightarrow C_{1}$. Moreover  
 	 $R\, \vdash_{\cal 
           C}\, S$, then
           $R'\, \vdash_{\cal C}\, S'$, because $R' \equiv R$ and  
         $S' \equiv S \wedge (C \Rightarrow  C_{1})$. Now, from 
         $R'\, \vdash_{\cal C}\, S'$ and the ${\cal C}$-satisfiability
         of $\Pi' R' \equiv \Pi R$, we deduce
         that $\Pi' S'$ is also ${\cal C}$-satisfiable. Therefore
the transformation step is allowed. 
 	 
 	2.   ${\cal G'} \subset {\cal G}$, so    
             	 $\Delta'; R', C'\, \vdash_{\cal UC}\, G'$ for all
             	 $\langle \Delta', C', G' \rangle \in {\cal G'}$ and 
 		${\cal M}_{{\cal G'}  R'} << {\cal M}_{{\cal G}  R}$.
		 
 	\item  If $G$ is atomic $G \equiv A$,  by hypothesis $\Delta; R, C\, 
 		\vlargo\, A$ has a proof of size  $l$, then by reason of the form 
of ${\cal UC}$, if $x_{1}, \ldots,x_{n}$ are new variables not free 
in  
 	 $\Delta$, $R$, $C$ neither in  $A$, then there is a variant of a 
formula from  $elab(\Delta)$,
 		$\forall x_{1}\ldots \forall x_{n}(G_1 \Rightarrow  A')$,
 		 with $A$ and $A'$ beginning with the same predicate symbol, such 
that   $\Delta; R, C\, 
 		\vlargo\, \exists x_{1}\ldots \exists x_{n}
 		((A' \approx A) \wedge G_{1})  (\dag)$
 		 has a proof of size less than $l$.
 		We transform  ${\cal S}$ in ${\cal S'}$ 
 		by means of the rule  {\em viii)}, using
 			$\forall x_{1}\ldots \forall x_{n}(G_1 \Rightarrow  A')$. 
        Assume now 
 		$R' \equiv R$. Since $S \equiv S'$ and $\Pi \equiv \Pi'$, the 
proof of 1. is immediate.
 		  
 2. Let $\langle  \Delta', C', G' \rangle \in {\cal G}'$,
    if  $\langle \Delta', C', G' \rangle \in {\cal G}$, then
 	$\Delta';R, C'\, \vdash_{\cal UC}\, G'$  by hypothesis
 	and therefore $\Delta'; R', C'\, \vdash_{\cal UC}\, G'$,
 	besides $\tau_{R'}(\Delta', C', G') =
 	 \tau_{R}(\Delta',  C', G')$.\\ 
 		 If  $\langle  \Delta', C', G' \rangle \notin {\cal 
 		G}$, then  $G' \equiv \exists x_{1}\ldots \exists x_{n}
 		((A' \approx A) \wedge G_{1})$, $C' \equiv C$
 		 and $\Delta' \equiv \Delta$. As we have noted in $(\dag)$,  
 	 $\Delta; R', C'\, 
 		\vlargo\, G'$
 		has a proof of size less than $l$.
	\end{itemize} 
\vspace{-2mm}
\hspace*{8.2mm} So 
 		 $\tau_{R'}(\Delta', C', G') < \tau_{R}(\Delta, C, G)$, and 2. 
is also proved in this case.
 \end{proof}

\end{document}